\documentclass[10pt,a4paper,aps,prd,showpacs,amsmath,amssymb,nofootinbib,showkeys,superscriptaddress,preprintnumbers]{revtex4-1}

\usepackage[german,english]{babel}   
\usepackage{hyperref}
\usepackage{bm}
\usepackage{bbm}
\usepackage{braket}
\usepackage{slashed}
\usepackage{multirow}
\usepackage{epsfig}
\usepackage{epstopdf}
\usepackage[mathlines]{lineno}
\usepackage{xcolor}

\allowdisplaybreaks

\DeclareMathOperator\arctanh{arctanh}
\DeclareMathOperator\arccot{arccot}
\newcommand{\ka}{\kappa}
\newcommand{\D}{\Delta}
\newcommand{\s}{\sigma}

\newcommand{\MS}{\overline{\rm MS}}

\begin{document}

\preprint{TUM-EFT 58/14}

\title{Effective field theories for van der Waals interactions}

\author{Nora Brambilla}
\affiliation{Physik-Department, Technische Universit\"at M\"unchen, James-Franck-Str. 1, 85748 Garching, Germany}
\affiliation{Institute for Advanced Study, Technische Universit\"at M\"unchen, Lichtenbergstrasse 2~a, 85748 Garching, Germany}
\author{Vladyslav Shtabovenko}
\author{Jaume Tarr\'us Castell\`a}
\author{Antonio Vairo}
\affiliation{Physik-Department, Technische Universit\"at M\"unchen, James-Franck-Str. 1, 85748 Garching, Germany}

\date{\today}

\begin{abstract}
Van der Waals interactions between two neutral but polarizable systems at a separation $R$ much larger than the typical size of the systems are at the core of a broad sweep of contemporary problems in settings ranging from atomic, molecular and condensed matter physics to strong interactions and gravity. In this paper, we reexamine the dispersive van der Waals interactions between two hydrogen atoms. 
The novelty of the analysis resides in the usage of nonrelativistic effective field theories of quantum electrodynamics. In this framework, the van der Waals potential acquires the meaning of a matching coefficient in an effective field theory, dubbed van der Waals effective field theory, suited to describe the low-energy dynamics of an atom pair. It may be computed systematically as a series in $R$ times some typical atomic scale and in the fine-structure constant $\alpha$. The van der Waals potential gets short-range contributions and radiative corrections, which we compute in dimensional regularization and renormalize here for the first time. Results are given in $d$ space-time dimensions. One can distinguish among different regimes depending on the relative size between $1/R$ and the typical atomic bound-state energy, which is of order $m\alpha^2$. Each regime is characterized by a specific hierarchy of scales and a corresponding tower of effective field theories. The short-distance regime is characterized by $1/R \gg m\alpha^2$ and the leading-order van der Waals potential is the London potential. We compute also next-to-next-to-next-to-leading order corrections. In the long-distance regime we have $1/R\ll m\alpha^2$. In this regime, the van der Waals potential contains contact terms, which are parametrically larger than the Casimir-Polder potential that describes the potential at large distances. In the effective field theory, the Casimir-Polder potential counts as a next-to-next-to-next-to-leading-order effect. In the intermediate-distance regime, $1/R\sim m\alpha^2$, a significantly  more complex potential is obtained. We compare this exact result with the two previous limiting cases. We conclude bz commenting on the van der Waals interactions in the hadronic case.
\end{abstract}


\maketitle

\section{Introduction}
Long-range electromagnetic forces between neutral particles in the absence of external electromagnetic fields have been studied for a long time, and are called {\em van der Waals interactions}. In this context, forces are of long range if they act at distances $R$ between the neutral particles much larger than the typical size of the particles. In the case of atoms, van der Waals forces act at distances $R \gg a_0$, where $a_0$ is the Bohr radius. 

Van der Waals interactions can be of different nature depending on whether they are generated by permanent dipoles (or higher multipole moments) or by instantaneously induced dipoles (or higher multipoles). In the latter case, the first studies were done by London in 1930~\cite{london} and included only the electrostatic Coulomb interaction, i.e., in a field theory language, only potential or Coulombic photons. London realized that systems without permanent dipole moments can still interact electromagnetically at second order in quantum-mechanical perturbation theory due to the mutually induced electric dipole moments. More precisely, they give rise to an attractive interaction that depends on the distance of the neutral particles as $1/R^6$: 
\begin{equation}
W_{\rm Lon} = -\frac{C_6}{R^6},
\label{london}
\end{equation}
which is known as the {\em London potential}.
The positive coefficient $C_6$ is a function of the instantaneous dipole moments of the interacting systems computed for all intermediate states and their energies. The computation of $C_6$ may be quite challenging for complex atomic and molecular systems. Furthermore, London related the strength of this interaction to the oscillator strengths of the system. Since the oscillator strengths are also related to the dispersion of light by the system, this type of van der Waals interactions are sometimes referred to as {\em dispersion forces}. Dispersion forces in this framework are therefore electromagnetic forces acting between well-separated neutral, unpolarized and unmagnetized atoms or bodies in the absence of any applied electromagnetic field.

To consider only potential photons is a good approximation as long as the interactions occur over small enough distances that the photon travel time is negligible. Casimir and Polder (CP) showed that retardation effects are important for the long-range regime, i.e., when $R$ is much larger than the typical time scale of the interacting particles~\cite{caspol}. They calculated a general form for the interaction from quantum mechanics using two-photon exchanges. Their result reproduces the London form at short distances, but at large distances, where retardation effects are important, the  van der Waals potential shows a different $R$ dependence:
\begin{equation}
W_{\rm CP} = - \frac{C_7}{R^7}\,,
\label{CP}  
\end{equation} 
where $C_7$ is a positive coefficient. The potential \eqref{CP} is also known as the {\em Casimir-Polder potential}. The results for the shorter distance regime, Eq.~\eqref{london}, and for the longer distance regime, Eq.~\eqref{CP}, were later reproduced using dispersive methods by Feinberg and Sucher~\cite{Feinberg:1970zz,Feinberg:1989ps}.

Dispersive forces are the weakest and at the same time the most persistent of all electromagnetic interactions. For this reason they play an important role across molecular physics, surface physics, colloid science, biology and even astrophysics. For the strong interactions, van der Waals potentials play a similar prominent dynamical role in systems made of at least two heavy quarks as they do in quantum electrodynamics (QED). The long-range interaction of small color dipoles [well realized in nature by heavy quarkonium states like the $J/\psi$ or $\Upsilon(1S)$] has been the subject of studies since the first years of quantum chromodynamics (QCD)~\cite{Bhanot:1979vb}. The interest in these systems is motivated by computing nuclear cross sections for quarkonium propagating in nuclear matter (relevant, e.g., for experiments at FAIR) or disentangling cold matter from deconfinement effects in heavy-ion collisions (relevant, e.g., for experiments at RHIC and LHC). The heavy quarkonium-nucleon scattering appears to be dominated by gluonic van der Waals interactions~\cite{Brodsky:1989jd,Luke:1992tm,Brodsky:1997gh,Beane:2014sda}. The quarkonium-quarkonium van der Waals interactions contain nontrivial information about low-energy QCD~\cite{Fujii:1999xn,Brambilla:2015rqa} and about quarkonium (chromo)polarizabilities~\cite{Voloshin:2007dx}. As in QED, the QCD van der Waals potential is attractive and in principle could lead to molecular-like bound states of heavy quarkonia with nuclei. This could possibly explain~\cite{Brambilla:2015rqa,Perevalova:2016dln}, for instance, the structure of the charmonium pentaquark observed at the LHCb experiment at CERN~\cite{Aaij:2015tga}. Van der Waals interactions are also prominent in studies of Feshbach resonances~\cite{Elhatisari:2013swa} and in the computation of quantum corrections to the gravitational potential between a pair of polarizable objects~\cite{Ford:2015wls}.

The broad interest for van der Waals interactions calls for a flexible, rigorous and systematic computational method that allows to properly define and efficiently compute van der Waals interactions in quantum field theory at any precision. Nonrelativistic effective field theories (EFTs)~\cite{Brambilla:2004jw} provide such a method. Recently there have been some studies of the electromagnetic van der Waals interactions in the framework of EFTs by Holstein~\cite{Holstein:2008fs}. In this paper, we will construct a complete set of effective field theories suited to compute the van der Waals interactions between two hydrogen atoms in different regimes. As we will see, our approach gives a clear definition and a computational framework for the potentials, without additional requirements~\cite{Sucher:2000yk}. The EFT approach developed here for the simpler case of QED will possibly provide a useful guideline for more complicated cases, like the study of van der Waals interactions in QCD that may also require dealing with nonperturbative effects.

The most simple polarizable neutral system is the hydrogen atom. The hydrogen atom is a nonrelativistic bound state characterized by three well-separated energy scales, which are the mass of the electron, $m$, the typical relative momentum, given by the inverse Bohr radius $1/a_0\sim m\alpha$, and the typical binding energy, which is of order $m\alpha^2$, where $\alpha = e^2/(4\pi) \approx 1/137 \ll 1$ is the fine-structure constant. These are usually referred to as  {\em hard}, {\em soft} and {\em ultrasoft scales} respectively. The presence of a set of well-separated energy scales makes the hydrogen atom a perfect system to apply nonrelativistic EFT techniques. Owing to their power counting, EFTs  significantly simplify bound-state calculations. Moreover, they are renormalizable (finite) order by order in the expansion parameters, which are $\alpha$ and ratios among the energy scales of the system. 
 
The EFT that follows from QED by integrating out the hard scale is nonrelativistic QED (NRQED)~\cite{Caswell:1985ui,Kinoshita:1995mt}. NRQED exploits the nonrelativistic nature of the electron and nucleus, but it does not yet take full advantage of the hierarchy of scales present in the system. The latter is achieved by potential NRQED (pNRQED)~\cite{Pineda:1997ie,Pineda:1998kn}, where both the hard and soft scales are integrated out. pNRQED provides a systematic description of the hydrogen atom derived from QED and a simpler and more efficient scheme for calculating all of the hydrogen properties in perturbation theory. In particular potentials appear in pNRQED as matching coefficients and the leading-order equation of motion of the nucleus-electron field is the Schr\"odinger equation.

When considering the van der Waals interactions between two hydrogen atoms the distance $R$ between them provides a new scale. As mentioned at the beginning, these interactions are well defined at a distance large enough that the internal structure of the atoms cannot be resolved, i.e., when $R$ is much larger than the Bohr radius: $R\gg a_0$.

The interplay between the scale $R$ and the typical time scale of the hydrogen atom, which is of order $1/(m\alpha^2)$, leads to different forms of the van der Waals interactions. There are three possible regimes: {\it (i)} the short-distance regime when $R\ll  1/m\alpha^2$, {\it (ii)} the long-distance regime when $R\gg 1/m\alpha^2$~\cite{Vairo:2000ia}, and {\it (iii)} the intermediate-distance regime when  $R\sim 1/m\alpha^2$. The aim of this paper is to work out an appropriate EFT and to compute the van der Waals potential between two hydrogen atoms for each of these physical situations. Results will be given in $d$ space-time dimensions and renormalized. The proper renormalization of the van der Waals interactions is one of the original contributions of the present work.
In the main body of the paper we will focus on atoms in $S$-wave states. More general results can be found in the appendices.

The paper is organized as follows. In Sec.~\ref{bipnrqed}, we summarize pNRQED. The van der Waals potentials are defined in the low-energy EFT of Sec.~\ref{vdweftsec}, which we call van der Waals EFT (WEFT). The short-, long- and intermediate-distance van der Waals potentials are computed in Secs.~\ref{srvdwi}, \ref{lrvdwi} and \ref{h3} respectively. Our summary and conclusions are in Sec.~\ref{conc}, where we also briefly discuss the relevance of the EFT framework for hadronic van der Waals interactions. The expressions for the loop integrals are given in Appendix~\ref{looi}. In Appendix~\ref{loodi}, we provide expressions for the dispersive van der Waals potentials for hydrogen atoms in any angular momentum state as well as the potentials generated by magnetic interactions. In Appendix~\ref{sumst}, we show some cases where the sum over the intermediate states can be performed explicitly. Finally, in Appendix~\ref{foutra} we list the Fourier transforms necessary to obtain the potentials in coordinate space.

\section{\texorpdfstring{\lowercase{p}}{p}NRQED} 
\label{bipnrqed}
Nonrelativistic bound states, such as the hydrogen atom, exhibit a hierarchy of well-separated scales. Integrating out the hard scale leads to NRQED. The EFT that results from integrating out the soft scale of order $1/a_0\sim m\alpha$ from NRQED is pNRQED\footnote{Note that it is also possible to define pNRQED as the EFT at the ultrasoft scale. The two definitions are equivalent in the one-atom sector, but differ in the two-atom one.}, which we briefly review in this section. The hydrogen atom is most suitably described in pNRQED.

The matching from QED to NRQED at one loop in the bilinear fermion sector was carried out using dimensional regularization in Ref.~\cite{Manohar:1997qy}. The NRQED Lagrangian density for two fermion species corresponding to electrons, $\psi$, and protons, $N$, up to $1/m$ corrections reads 
\begin{equation}
  \mathcal{L}_{\text{NRQED}}^{\text{2-fermion}} =
  \psi^{\dagger}\left[i D^0+\frac{\boldsymbol{D}^2}{2m} - c_F \frac{\boldsymbol{\sigma}\cdot e\boldsymbol{B}}{2m} \right]\psi
  +N^{\dagger}iD^0N -\frac{1}{4}F^{\mu\nu}F_{\mu\nu}\,,
\label{Lag-NRQED}  
\end{equation}
where $m$ is the electron mass and $-e$, with $e >0$, is the electron charge. Here $F^{\mu \nu} =\partial^\mu A^\nu - \partial^\nu A^\mu$ is the electromagnetic field-strength tensor and $A^\mu$ is the photon field. Throughout this paper we use bold font to specify Cartesian components of 4-vectors. The covariant derivatives acting on the electron field are defined as $iD^0=i\partial^0+ e A^0$, $i\bm{D}=i\bm{\nabla} - e \bm{A}$, and the one acting on the proton field is defined as $iD^0=i\partial^0 - e A^0$. The proton mass $M$ is taken to be much larger than the electron mass, and therefore operators proportional to powers of $1/M$ are beyond the precision we are aiming at and will be neglected. The matching coefficient $c_F$ is one half of the electron magnetic moment; at order $\alpha$ it reads 
\begin{equation}
c_F = 1 + \frac{\alpha}{2\pi}\,. 
\end{equation}
Requiring that the electric charge in Eq.~\eqref{Lag-NRQED} is the one measured in low-energy experiments (e.g., Thomson scattering) guarantees that the matching coefficient of the operator $-F^{\mu\nu}F_{\mu\nu}/4$ remains one to all orders~\cite{Pineda:1998kj}. The matching coefficient of the kinetic energy operator is also constrained to be one to all orders by reparametrization/Poincar\'e invariance.

For the purpose of renormalizing the van der Waals interactions, as we will see, we need to add to Eq.~\eqref{Lag-NRQED} four-electron operators. The four-electron operators of dimension six are 
\begin{equation}
  \mathcal{L}_{\text{NRQED}}^{\text{4-fermion}} = 
  \frac{d_s}{m^2}\left(\psi^{\dagger}\psi\right)^2+\frac{d_v}{m^2}\left(\psi^{\dagger}\boldsymbol{\sigma}\psi\right)^2.
\label{4fop}
\end{equation}
The matching coefficients at one loop in the $\MS$ renormalization scheme read~\cite{Pineda:1998kj,Brambilla:2005yk}
\begin{equation}
d_s=\alpha^2\left[\log\left(\frac{m^2}{\nu^2}\right)-\frac{2}{3}\right], \qquad d_v=\alpha^2\,,
\label{dsdv}
\end{equation}
where $\nu$ is the renormalization scale. The divergence in $d_s$ is of infrared origin and cancels in physical observables against ultraviolet divergences coming from low-energy modes.

To obtain pNRQED we integrate out electrons and photons with a soft momentum. The soft scale is given by the typical relative momentum between the electron and the proton, which is of size $m\alpha$. Since the energies and momenta that we are integrating out are of order $m\alpha$, we can use static propagators for the electron field to perform the matching. This allows to match NRQED to pNRQED at any given order in $1/m$ and $\alpha$.

It is convenient to introduce the center-of-mass coordinate $\bm{X} \approx \bm{x}_p(1 + \mathcal{O}(m/M))$ and the electron-proton distance $\bm{x}= \bm{x}_e - \bm{x}_p$, where $\bm{x}_e$ and $\bm{x}_p$ are the coordinates of the electron and proton respectively. Variations in the center-of-mass coordinate due to recoiling of the atom against low-energy photons are of the order of the  inverse of the ultrasoft scale, and hence much smaller than the typical magnitude of $\bm{x}$, which is of the order of the inverse of the soft scale. The dynamical degrees of freedom of pNRQED are: the field $S(t,\bm{x},\bm{X})$, which is invariant under $U(1)$ gauge transformations, and encodes the proton and electron fields, and photons $A_\mu (t,\bm{X})$. The photon fields have been multipole expanded in $\bm{x}$ to guarantee that they are ultrasoft. The Lagrangian also contains potential terms, that is, terms that are independent of time and nonlocal in space, which naturally arise in the matching of NRQED to pNRQED. 

Showing only operators that are relevant for our study, the pNRQED Lagrangian density at $\mathcal{O}(\bm{x},1/m)$ reads~\cite{Pineda:1997ie}
\begin{equation}
  \mathcal{L}^{\text{1-atom}}_{\text{pNRQED}} = \int d^3\bm{x} \, S^{\dagger}(t,\bm{x},\bm{X})
  \left[i\partial_0+\frac{\bm{\nabla_{\bm{x}}^2}}{2m}+\frac{\alpha}{|\bm{x}|} - \bm{x}\cdot e\bm{E} - \frac{\bm{\mu}\cdot e\bm{B}}{2m}\right]S(t,\bm{x},\bm{X})
  -\frac{1}{4}F^{\mu\nu}F_{\mu\nu}\,,
\label{pnrqedl}
\end{equation}
where the total magnetic moment of the electron is defined as $\bm{\mu}=\bm{L}+2c_F\bm{S}$, where $\bm{S}=\bm{\sigma}/2$ and $\bm{L}=-i (\bm{x}\times\bm{\nabla}_{\bm{x}})$ are the electron spin and orbital angular momentum operators. The operators $\bm{x}\cdot e \bm{E}$ and $\bm{\mu}\cdot e \bm{B}/(2m)$ are the electric and magnetic dipoles respectively. All electromagnetic fields in the Lagrangian density are located at $(t,\bm{X})$. The size of each term in Eq.~\eqref{pnrqedl} can be evaluated as follows. Each relative momentum $-i\bm{\nabla}_{\bm{x}}$ and inverse relative coordinate $1/x$ have a size of $m\alpha$. Time derivatives acting on the atom field $S(t,\bm{x},\bm{X})$ have a size of $m\alpha^2$. Each ultrasoft photon field and derivatives acting on it are of order $m\alpha^2$, which leads to $\bm{E}\sim\bm{B}\sim m^2\alpha^4$. In the two-atom sector the four-electron operators of Eq.~\eqref{4fop} induce at leading order in the multipole expansion a contact interaction between $S$ fields, which is 
\begin{equation}
  \mathcal{L}^{\text{2-atom}}_{\text{pNRQED}} =
  \int d^3\bm{x}_1 d^3\bm{x}_2 \,\left[ 
  \frac{d_s}{m^2}S^{\dagger}S(t,\bm{x}_1,\bm{X}) \, S^{\dagger}S(t,\bm{x}_2,\bm{X})
  +\frac{d_v}{m^2}S^{\dagger}\boldsymbol{\sigma}S(t,\bm{x}_1,\bm{X}) \cdot S^{\dagger}\boldsymbol{\sigma} S(t,\bm{x}_2,\bm{X})\right].
\label{2sci}
\end{equation}

\section{Van der Waals EFT}
\label{vdweftsec}
The energy scale, $Q$, at which the dynamics of the two hydrogen atoms happens is of order $\bm{k}^2/M$, where $\bm{k}$ is the typical momentum transfer between the atoms and $M$ is the proton mass, which here is a good approximation of the total mass of the electron-nucleus system. If we restrict ourselves to interactions over distances larger than the Bohr radius of the atoms, $R \gg a_0$, then the typical momentum $\bm{k}$ is much smaller than $ m \alpha$ and the energy scale of the two-atom dynamics is much smaller than the ultrasoft scale. At an energy scale $Q$ photons of higher energy are not resolved and their effect is taken into account by potential terms. We are going to refer to these latter terms as {\em van der Waals potentials} and to the EFT that lives at the scale $Q$ and describes the dynamics of hydrogen atoms interacting through them as {\em van der Waals EFT} (WEFT). For the QCD equivalent of this EFT see Ref.~\cite{Brambilla:2015rqa}. 

The degrees of freedom of WEFT are $U(1)$ singlet fields, $S_n$, describing atoms with quantum numbers $n$, and low-energy photons carrying momentum and energy of order $Q$. At energies much below the typical binding energy $E_n \sim m\alpha^2$, the different atomic states are frozen, for photons are not energetic enough to excite them, and have to be considered as independent fields. Hence, in the absence of external interactions, the fields $S_n$ are plane waves of energy $E_n$, where $E_n$ is the binding energy of the state $|n\rangle$ solution of the Schr\"odinger equation for the hydrogen atom, at leading order $E_n = -m\alpha^2/(2n^2)$. The Lagrangian is built by coupling these states to electromagnetic fields and to each other. Since $S_n$ are nonrelativistic fields, the couplings to the electromagnetic fields are better expressed in terms of the electric field, $\bm{E}$, and magnetic field, $\bm{B}$. The couplings between $S_n$ are expressed in the Lagrangian by potentials.

In the one-atom sector, in going from QED to WEFT we integrate out the scales $m$, $m\alpha$ and $m\alpha^2$, and thus one should be able to organize the WEFT Lagrangian as a series in the ratios \begin{equation}
\frac{m\alpha}{m},\,\, \frac{m\alpha^2}{m\alpha},\,\, \frac{Q}{m\alpha^2}\,.
\label{ratios}
\end{equation}
The scaling of the singlet field is $S_n\sim Q^{3/2}$. Space and time derivatives acting on the electromagnetic field $\bm{A}$ and the field itself are of order $Q$. Temporal derivatives acting on the singlet field scale like $Q$ but space derivatives scale like $\sqrt{MQ}$. 

The Lagrangian of WEFT in the one-atom sector reads
\begin{equation}
  L^{\text{1-atom}}_{\text{WEFT}} = \int d^3\bm{X}\,\sum_{n} S_n^{\dagger}(t,\,\bm{X})
  \left[i\partial_0-E_{n}+\frac{\bm{\nabla_{\bm{X}}^2}}{2M}+2\pi\alpha^{ij}_{n}\bm{E}_i\bm{E}_j+2\pi \beta^{ij}_{n} \bm{B}_i\bm{B}_j
    -\frac{\langle n|\bm{\mu}|n\rangle \cdot e \bm{B}}{2m} + \dots \right]S_n(t,\,\bm{X}),
\label{aeft1sl}
\end{equation}
where the dots stand for multipolar couplings and higher-order terms. Note that, in general, $E_n$ is a matrix with an imaginary part accounting for transitions between energy levels. Because the mixing of states is irrelevant for this paper, we have neglected it in Eq.~\eqref{aeft1sl}.

\begin{figure}[ht]
\centering{\includegraphics[width=0.35\textwidth]{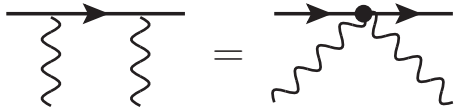}}
\caption{Tree-level matching of the couplings of the hydrogen atom with external radiation fields. On the left-hand side we have the pNRQED diagram (photons couple to atoms either via electric or magnetic dipoles; see Eq.~\eqref{pnrqedl}) and on the right-hand side the WEFT one.}
\label{h2tree}
\end{figure}

In the one-atom sector the matching is performed by equating Green's functions calculated in WEFT to the ones calculated in pNRQED in the limit of the external energies being much smaller than the bound-state energies. The matchings of the kinetic operator and of the coupling to the magnetic field are trivial. The two electromagnetic field operators are obtained by matching the right-hand side of Fig.~\ref{h2tree} with the left-hand side:
\begin{equation}
\alpha^{ij}_{n}=\frac{1}{2\pi}\sideset{}{'}\sum_m \frac{p_E(n,m)^{ij}}{\D E_{nm}}\,,\qquad
\beta^{ij}_{n}=\frac{1}{2\pi }\sideset{}{'}\sum_m\frac{p_B(n,m)^{ij}}{\D E_{nm}}\,.
\label{pol0}
\end{equation}
A prime in the sum sign, here and in the following, signifies that it runs over all values of the index/indices except the one/ones labeling the incoming energy [in Eq.~\eqref{pol0} this is $n$]. 
The sum is a shorthand notation that also encompasses the integral over the continuum states. Moreover, we have used the following notations: $\D E_{nm}=E_n-E_m$, and 
\begin{eqnarray}
p_E(n,m)^{ij}&=&e^2\langle n|\bm{x}^i|m\rangle \langle m|\bm{x}^j|n\rangle \,,\\
p_B(n,m)^{ij}&=&\frac{e^2}{4m^2}\langle n|\bm{\mu}^i|m\rangle \langle m|\bm{\mu}^j|n\rangle\,.
\end{eqnarray}
The couplings $\alpha^{ij}_{n}$ and $\beta^{ij}_{n}$ are called the {\em static electric and magnetic polarizability} tensors~\cite{pSchwer}. 

For the hydrogen atom the dipole moment $\langle n|\bm{x}|n\rangle$ vanishes due to parity, however, higher multipole moments are allowed. For instance, for states with $L \ge 2$ the quadrupole moment does not vanish. In the present work, we will mostly focus on the study of the dispersive van der Waals interactions between $S$-wave states in which all multipole moments vanish. In this case the electric polarizability takes a scalar form $\alpha^{ij}_{n}=\alpha_{n}\delta^{ij}$ and the magnetic dipole is given by the spin only: $\langle n | \bm{\mu} |n\rangle =2 c_F \langle n|\bm{S}|n\rangle$.
 
If both electric and magnetic polarizabilities take a scalar form, then the corresponding part of the Lagrangian simplifies to $2\pi\alpha_{n} \bm{E}^2 + 2\pi\beta_{n}\bm{B}^2$, which can be found in studies of neutral particles interacting with electromagnetic fields (see, e.g., Refs.~\cite{Holstein:1999uu,Kaplan:2005es,Griesshammer:2004yn}). However, for the hydrogen atom the magnetic polarizability never takes the scalar form since the hydrogen atom has a permanent magnetic dipole due to the spin of the electron.

In the two-atom sector the WEFT Lagrangian contains potential interactions between atoms:
\begin{equation}
L^{\text{2-atom}}_{\text{WEFT}} = -\int d^3 \bm{X}_1 \, d^3 \bm{X}_2 \, 
\sum_{n_i,n_j} S_{n_i}^{\dagger}(t,\bm{X}_1)S_{n_i}(t,\bm{X}_1)\,W_{n_i,\,n_j}(\bm{X}_1-\bm{X}_2)\,S_{n_j}^{\dagger}(t,\bm{X}_2)S_{n_j}(t,\bm{X}_2) \,.
\label{2sp2h}
\end{equation}
The potential $W_{n_i,\,n_j}$ corresponds to the van der Waals potential between two atoms in a $|n_i\rangle$ and $|n_j\rangle$ state respectively. Note that, in general, the theory will also contain potential interactions that change the state of the hydrogen atoms involved, and couplings of the two atoms with photons of energy $Q$. However, neither of them is going to contribute to the atom-atom interaction at the accuracy of this work. In the following, we will often omit the indication of the states and denote the van der Waals potential simply by $W$ and its Fourier transform in momentum space by~$\widetilde{W}$.

A new physical scale appears in the van der Waals potentials of Eq.~\eqref{2sp2h}: the distance ${\bf R} = \bm{X_1}-\bm{X_2}$ between the two interacting atoms, whose conjugate variable is the momentum transfer between the atoms, which we denote by~$\bm{k}$. The potential in the two-atom sector can be expressed as an expansion in the ratios of scales of Eq.~\eqref{ratios} as well as an expansion in $(m\alpha^2\, R)$ for the short-distance regime ($R \ll 1/(m \alpha^2)$) and in $1/(m\alpha^2 \, R)$ for the long-distance regime ($R \gg 1/(m \alpha^2)$). In the intermediate-distance regime, where $1/R \sim m \alpha^2$,  we cannot expand $R$ with respect to $1/(m\alpha^2)$. In the following sections, we will to obtain $W$ for these different regimes.

\section{Short-range Van der Waals interactions}
\label{srvdwi}
\begin{figure}[ht]
\begin{tabular}{cc}
  \includegraphics[width=.40\textwidth]{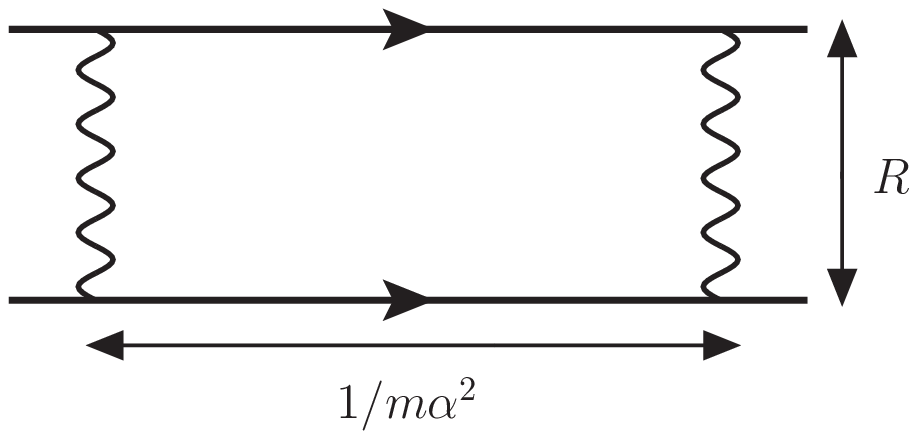}  & \hspace{2cm}  \includegraphics[width=.3\textwidth]{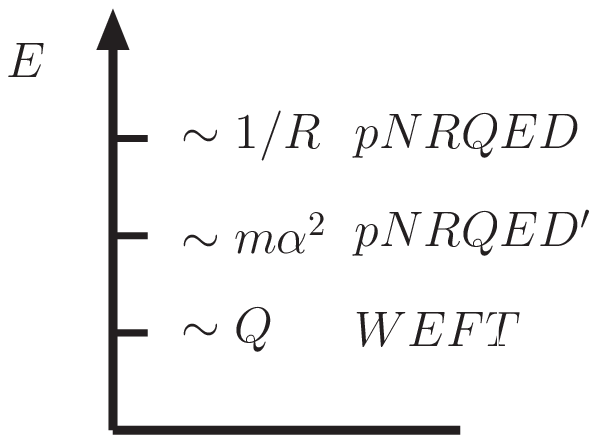}
\end{tabular}
\caption{Left panel: Sketch of the physical picture of the van der Waals interactions in the short-range regime. The distance between the atoms is much smaller than the typical time scale of the hydrogen atom but much larger than the Bohr radius. Right panel: Hierarchy of scales and the corresponding EFTs in the short-distance regime.}
\label{h1sk}
\end{figure}

In this section, we study the van der Waals interactions in the short-distance regime. The physical situation is sketched in the left-hand panel of Fig.~\ref{h1sk}. The characteristic of this regime is that the distance between the atoms is much smaller than the time scale between the emission of the two photons, which is of the order of the inverse of the ultrasoft scale: $R \ll 1/(m \alpha^2)$. On the other hand, the distance between the atoms is much larger than the Bohr radius, namely the typical size of a hydrogen atom: $R \gg a_0$. Photons exchanged between the atoms carry a typical momentum, $\bm{k}$, that is of the order of the inverse of the distance between the atoms or $m \alpha^2$ or smaller. Finally, the energy scale of the atoms is set by the kinetic energy $Q \sim \bm{k}^2/M$. Since $M$ provides a very strong suppression, the dynamics of the atoms occurs at a scale much smaller than any of the previous ones. In practice, for the purpose of computing the van der Waals potential we may consider the atoms as static. In the right-hand panel of Fig.~\ref{h1sk} we have plotted the hierarchy of scales involved together with the corresponding EFTs.

In the following sections we will provide the details of the matching between the hierarchy of EFTs in the right-hand panel of Fig.~\ref{h1sk}. In Sec.~\ref{h1mpnrqed} we integrate out photons carrying momentum of order $1/R$ and obtain pNRQED$^{\prime}$. In the one-atom sector, pNRQED$^{\prime}$ is similar to the theory described in Sec.~\ref{bipnrqed}, but in the two-atom sector new potentials appear. The van der Waals interactions are obtained in Sec.~\ref{h1vdeft} by integrating out photons with energy and momentum of order $m\alpha^2$ and virtual atomic states, whose energy is also of order $m\alpha^2$, and matching pNRQED$^{\prime}$ to WEFT.

In order to make the counting homogeneous, it is convenient to assign a specific size to $1/R$ in terms of $m$ and $\alpha$. A natural choice, given the scale hierarchy in the short-distance regime, is to take $1/R \sim m \alpha \sqrt{\alpha}$. From this assignment it follows that $ (m\alpha^2\, R) \sim \sqrt{\alpha}$. In the short-distance regime, we aim at computing the van der Waals interactions, $W$ (in coordinate space), up to order $m \alpha^6\sqrt{\alpha}$.

\subsection{Matching pNRQED\texorpdfstring{$^{\prime}$}{'}}
\label{h1mpnrqed}
To obtain pNRQED$^{\prime}$ we have to integrate out photons whose momentum and energy scale like $1/R$. The one-atom sector does not change in matching pNRQED to pNRQED$^{\prime}$ and is given by Eq.~\eqref{pnrqedl}, whereas new potential interactions, $V$, arise in the two-atom sector:
\begin{equation}
  L^{\text{2-atom}}_{\text{pNRQED'}} =
  -\int d^3\bm{X}_1d^3\bm{X}_2 \, d^3\bm{x}_1d^3\bm{x}_2 \, S^{\dagger}S(t,\bm{x}_1,\bm{X}_1)\,V(\bm{X}_1-\bm{X}_2)\,S^{\dagger}S(t,\bm{x}_2,\bm{X}_2)\,.
\label{pnrqed2s}
\end{equation}

\begin{figure}[ht]
\centering{\includegraphics[height=2.3cm]{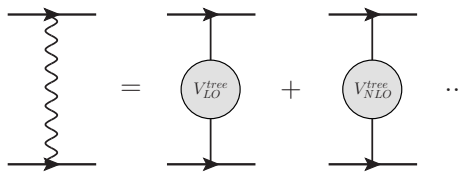}}
\caption{Tree-level matching of the two hydrogen atom potentials of pNRQED$^{\prime}$. The pNRQED diagram (the photon couples with electric or magnetic dipoles) is on the left-hand side and the pNRQED$^{\prime}$ ones are on the right-hand side .}
\label{h1tree}
\end{figure}

The matching of pNRQED to pNRQED$^{\prime}$ in the two-atom sector is given at tree level by the exchange of one photon (Fig.~\ref{h1tree}) and (at the order we are interested in) the contact interaction of Eq.~\eqref{2sci}. First, we consider the one-photon exchange diagram. Since the energy of the atoms is the smallest scale in the problem, the photon propagator can be expanded in it. Hence, the momentum of the photon scales with the only scale in the diagram, which is $1/R \sim m \alpha \sqrt{\alpha}$.

The leading-order (LO) contribution to Eq.~\eqref{pnrqed2s} is an electric dipole exchange. This is of $\mathcal{O}\left(1/(m^2\alpha)\right)$ in momentum space and reads
\begin{equation}
  \widetilde{V}^{\text{tree}}_{LO,\,E} = -e^2 \frac{\bm{x}_1\cdot\bm{k}\,\bm{x}_2\cdot\bm{k}}{\bm{k}^2}\,,
  \label{loeep}
\end{equation}
where $\bm{x}_1$ and $\bm{x}_2$ are the electron-proton distances in the two atoms. Using the Fourier transforms of Appendix~\ref{foutra} we can obtain the dipole potential in position space with its characteristic $R^{-3}$ dependence. In coordinate space the potential is of order $m \alpha^3\sqrt{\alpha}$.

At next-to-leading order (NLO) [$\mathcal{O}\left(1/m^2\right)$ in momentum space] we obtain
\begin{equation}
  \widetilde{V}^{\text{tree}}_{NLO,\,E} = e^2 \frac{\bm{v}_1\cdot\bm{k}\,\bm{v}_2\cdot\bm{k}-\bm{v}_1\cdot \bm{v}_2\bm{k}^2}{\bm{k}^4}\,.
  \label{nloeep}
\end{equation}
The NLO term is proportional to $\bm{v} = -i[\bm{x},\hat{h}_0]$, where $\hat{h}_0 = -\bm{\nabla}^2_{\bm{x}}/(2m)-\alpha/|\bm{x}|$. This dependence arises from the fact that the NLO contribution is proportional to the square of the energy carried by the photon. Using the equations of motion for the $S$ field it can be shown that\footnote{By means of Eq.~\eqref{eveq} the subleading term in the expansion of the one-photon exchange can be identified with a pNRQED potential. In principle, one could absorb both energy factors in either singlet pair or one factor per singlet pair. These choices are related by partial integration and correspond to different operator basis. In Eq.~\eqref{nloeep} we have chosen to absorb one energy factor for each singlet pair as it leads to a simpler one-loop matching calculation.}
\begin{equation}
  k_0S^{\dagger}\bm{x}S = i\left(\partial_0S^{\dagger}\bm{x}S+S^{\dagger}\bm{x}\partial_0S\right) =
  -\left(S^{\dagger}\hat{h}_0\bm{x}S-S^{\dagger}\bm{x}\hat{h}_0S\right) = S^{\dagger}[\bm{x},\hat{h}_0]S = iS^{\dagger}\bm{v}S\,.
\label{eveq}
\end{equation}
In coordinate space the NLO potential is of order $m \alpha^4\sqrt{\alpha}$.

The tree-level diagram on the left-hand side of Fig.~\ref{h1tree} may also be understood as an exchange between two magnetic dipoles or an electric dipole and a magnetic dipole. For the case of two magnetic dipoles we have
\begin{eqnarray}
  \widetilde{V}^{\text{tree}}_{LO,\,B}&=&\frac{e^2}{4m^2}\frac{1}{\bm{k}^2}\left(\bm{k}^2\bm{\mu}_1\cdot\bm{\mu}_2-\bm{\mu}_1\cdot\bm{k}\,\bm{\mu}_2\cdot\bm{k}\right),
  \label{bbp}
\end{eqnarray}
where the subindices on $\bm{\mu}$ and $\bm{S}$ identify the atom. This is a contribution of order $\alpha/m^2$ in momentum space (of order $m \alpha^5\sqrt{\alpha}$ in coordinate space).

Unlike the two former cases, the electric-magnetic dipole interaction is proportional to $k^0$ at leading order. We use Eq.~\eqref{eveq} to convert the $k^0$ factor into a time derivative of $S$ and write 
\begin{eqnarray}
  \widetilde{V}^{\text{tree}}_{LO,\,M}&=&i\frac{e^2}{2m}\frac{\bm{k}}{\bm{k}^2}\cdot\left(\bm{v}_1\times \bm{\mu}_2-\bm{\mu}_1\times\bm{v}_2\right).
  \label{ebp}
\end{eqnarray}
This contribution is of order $\sqrt{\alpha}/m^2$ in momentum space (of order $m \alpha^5$ in coordinate space).

The matching of the contact interaction of Eq.~\eqref{2sci} is trivial since it is independent of the momentum:
\begin{equation}
\widetilde{V}^{\text{cont.}} = - \frac{d_s}{m^2}-\frac{4d_v}{m^2}\bm{S}_1\cdot \bm{S}_2\,.
\label{ctr}
\end{equation}
This contribution is of $\mathcal{O}\left(\alpha^2/m^2\right)$ in momentum space ($\mathcal{O}\left(m \alpha^6\sqrt{\alpha}\right)$ in coordinate space).

Note that, by considering the power counting only, further subleading contributions from electric dipole, magnetic dipole and electric-magnetic dipole tree-level exchanges could, in principle, be of order $\alpha^2/m^2$ or larger in momentum space. Nevertheless, these terms do not eventually contribute to the van der Waals interactions. In fact the characteristic feature of dispersive van der Waals interactions is that they appear at the one-loop level. The reason is that the expectation value of the electric dipole for eigenstates of the hydrogen atom vanishes due to parity, as well as higher multipole moments for $S$-wave states. Moreover all contributions proportional to the energy transfer, like higher-order electric and magnetic dipole potentials, and electric-magnetic dipole potentials  vanish once evaluated on static initial- and final-state atoms with the same quantum numbers. In the case of hydrogen atoms, the only tree-level exchange that contributes to the the atom-atom interaction is the exchange of two magnetic dipoles: it does not depend on the energy transfer and the hydrogen atom has a permanent magnetic dipole generated by the spin of the electron. In addition, for nonzero orbital angular momentum, there is also an orbital contribution to the magnetic dipole. As a result, out of all potentials generated by a tree-level photon exchange, only $\widetilde{V}^{\text{tree}}_{LO,\,B}$ gives a nonvanishing contribution to the hydrogen-hydrogen interaction.

The situation changes at the loop level due to virtual intermediate states. Loops involving electric dipoles do not necessarily vanish because they are proportional to $\langle n|\bm{x}|m\rangle$ with $|m\rangle$ being an intermediate state different from $|n\rangle$. The intermediate states are the solutions of the Schr\"odinger equation with Hamiltonian $\hat{h}_0$ for the hydrogen atom; $n$ collects all discrete and continuum quantum numbers necessary to label the spectrum of $\hat{h}_0$. To perform the one-loop matching is, therefore, of paramount importance to compute the van der Waals interactions.

\begin{figure}[ht]
\centering{\includegraphics[height=2.3cm]{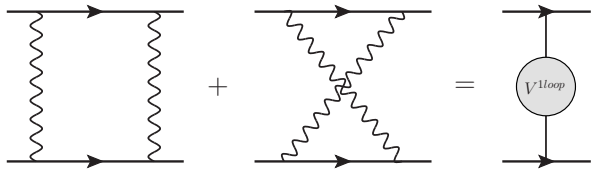}} 
\caption{One-loop matching of the two-atom potential of pNRQED$^{\prime}$. The pNRQED diagrams are on the left-hand side and the pNRQED$^{\prime}$ one is on the right-hand side.}
\label{h1m1}
\end{figure}

We proceed to analyze the one-loop contributions to the two-atom potential of Eq.~\eqref{pnrqed2s} up to $\mathcal{O}\left(\alpha^2/m^2\right)$ in momentum space. The scheme of the one-loop matching can be found in Fig.~\ref{h1m1}, with pNRQED diagrams on the left-hand side matching the pNRQED$^{\prime}$ diagram on the right-hand side. In the loop, we integrate over a photon momentum $q^\mu$ such that $q^0\sim|\bm{q}|\sim 1/R$. The one-loop diagrams give rise to new terms for the potential of Eq.~\eqref{pnrqed2s} starting at $\mathcal{O}(\alpha/m^2)$ in momentum space. Subleading contributions are suppressed by powers of $\sqrt{\alpha}$, and thus the first three terms are needed to reach an accuracy of $\mathcal{O}(\alpha^2/m^2)$. Pinch singularities cancel against the two dipole potential exchanges in pNRQED$^{\prime}$~\cite{Brambilla:2004jw}.

In these new potential terms powers of the energy gap factors $\D E_{nm}$ appear in the numerator, however the matching coefficients of pNRQED$^{\prime}$ cannot depend on a specific state. This dependence on $\D E_{nm}$ is fictitious and can be eliminated by using the results of Appendix~\ref{sumst} to perform the sums over the intermediate states. In the case that all the vertices are electric dipole couplings, it turns out that after summing over the intermediate states the LO and NLO contributions vanish. Only the next-to-next-to-leading-order (N$^2$LO) term survives:
\begin{equation}
  \widetilde{V}^{\text{1loop}}_{N^2LO,\,E}  = -\frac{2\pi^2\alpha^2}{3m^2}(d-2)(4d-9) A_{3/2}(\bm{k}^2) \,,
  \label{h1helo} 
\end{equation}
where $A_{3/2}$ is a loop integral defined in Appendix~\ref{looi} and $d$ is the space-time dimension. The expression in Eq.~\eqref{h1helo} is, indeed, independent of the initial and final states considered. The potential $\widetilde{V}^{\text{1loop}}_{N^2LO,\,E}$ is of order $\alpha^2/m^2$, whereas the corresponding expression in coordinate space, $V^{\text{1loop}}_{N^2LO,\,E}$, is of order $m \alpha^6\sqrt{\alpha}$. Analogous matching contributions can be obtained by replacing two or four of the electric dipole couplings by magnetic dipole ones. These are suppressed by $\alpha$ and $\alpha^3$ respectively with respect to the four electric dipole interaction computed in Eq.~\eqref{h1helo}, and are, therefore, beyond the precision we are aiming at. Nevertheless the corresponding expressions are given in Appendix~\ref{loodi}.

\subsection{Matching WEFT}
\label{h1vdeft}
\begin{figure}[ht]
\centering{\includegraphics[width=1.0\textwidth]{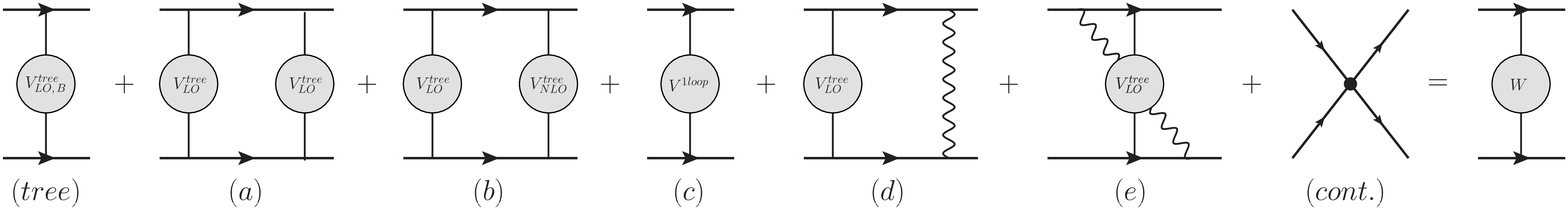}}
\caption{Matching of the van der Waals potential between two hydrogen atoms. The pNRQED$^{\prime}$ diagrams are on left-hand side and the WEFT one is on the right-hand side. The symmetric diagrams of {\it (b)}, {\it (d)} and {\it (e)} have not been displayed, but are understood.}
\label{h1m2}
\end{figure}

The last remaining step to obtain the van der Waals potential in the short-distance regime consists in integrating out ultrasoft photons with energy and momentum of order $m\alpha^2$ and virtual atomic states, whose energy is also of order $m\alpha^2$. This is done by matching the two-atom sector of pNRQED$^{\prime}$ from the previous section to WEFT. The relevant contributions to the van der Waals potential defined in Eq.~\eqref{2sp2h} can be found on the left-hand side of Fig.~\ref{h1m2}. The tree-level contribution is the magnetic dipole potential of Eq.~\eqref{bbp}. There are four one-loop diagrams to be considered: diagram {\it (a)} with two LO electric dipole potential exchanges, which is of $\mathcal{O}(\sqrt{\alpha}/m^2)$ ($m \alpha^5$ in coordinate space), diagram {\it (b)} with one LO and one NLO electric dipole potential, which is of $\mathcal{O}(\alpha\sqrt{\alpha}/m^2)$ ($m \alpha^6$ in coordinate space), and diagrams {\it (d)} and {\it (e)} with one LO electric dipole potential and one ultrasoft photon, which are of $\mathcal{O}(\alpha^2/m^2)$ ($m \alpha^6\sqrt{\alpha}$ in coordinate space). Of the same order as the latter are also diagram {\it (c)}, which is the potential of Eq.~\eqref{h1helo}, and the last diagram, which is the contact term of Eq.~\eqref{ctr}. Diagrams involving more than two potential exchanges, either vanish because of parity or give contributions beyond our accuracy. Diagrams consisting of the exchange of two ultrasoft photons contribute at order $\alpha^2/m^2$ in momentum space but only at order $m \alpha^8$ in coordinate space and are therefore also beyond our accuracy for the computation of $W$. 

The contribution from diagram {\it (a)} reads
\begin{equation}
\widetilde{W}^{(a)}_{E}=\frac{\bm{k}^4 A_2(\bm{k}^2)}{4}\sideset{}{'}\sum_{m_1,m_2}\frac{p_E(n_1,m_1)p_E(n_2,m_2)}{\Delta E_{n_1m_1}+\Delta E_{n_2m_2}}\,,
\label{2lopot}
\end{equation}
where $n_1$ and $n_2$ label the hydrogen atom states and $A_{2}$ is a loop integral defined in Appendix~\ref{looi}. We have used that for $S$-wave states, due to rotational symmetry, $p_E(n,m)^{ij}=p_E(n,m)\delta^{ij}$ (where a sum over all degenerate intermediate Coulomb states is understood).

The contribution from diagram {\it (b)} reads
\begin{equation}
\widetilde{W}^{(b)}_{E}=\bm{k}^2A_2(\bm{k}^2)\sideset{}{'}\sum_{m_1,m_2}p_E(n_1,m_1)p_E(n_2,m_2)\frac{\D E_{n_1m_1}\D E_{n_2m_2}}{\D E_{n_1m_1}+\D E_{n_2m_2}}\,,
\label{lonlopot}
\end{equation}
where we have used $\langle n|\bm{v}|m\rangle =  i\D E_{nm} \langle n|\bm{x}|m\rangle$.

Diagrams {\it (d)} and {\it (e)} involve one ultrasoft photon. Their contribution reads
\begin{equation}
  \widetilde{W}^{(d+e)}_{E} = \frac{4(d-2)}{d-1}\sideset{}{'}\sum_{m_1,m_2}p_E(n_1,m_1)p_E(n_2,m_2)\frac{\D E_{n_1m_1}\D E_{n_2m_2}}{\D E^2_{n_1m_1}-\D E^2_{n_2m_2}}
  \left[\D E_{n_1m_1}J(\D E_{n_1m_1}) - \D E_{n_2m_2} J(\D E_{n_2m_2})\right],
  \label{deh1}
\end{equation} 
where $J(\D E_{nm})$ is a loop integral that can be found in Appendix~\ref{looi} and $d$ is the space-time dimension. 

Finally, adding the contributions from the left-hand side of Fig.~\ref{h1m2}, we obtain all terms relevant to compute the van der Waals potential, $W$, up to order $m \alpha^6\sqrt{\alpha}$. In momentum space they read
\begin{eqnarray}
  \widetilde{W}^{(0)}&=&\widetilde{W}^{(a)}_{E}\,,
  \label{loh1ep} \\
  \widetilde{W}^{(1/2)}&=&\langle n_1,n_2|\widetilde{V}^{\text{tree}}_{LO,\,B}|n_1,n_2\rangle \,,
  \label{treeh1bp}\\
  \widetilde{W}^{(1)}&=&\widetilde{W}^{(b)}_{E}\,,
  \label{nloh1ep}\\
  \widetilde{W}^{(3/2)}&=&\widetilde{W}^{(d+e)}_{E}+\langle n_1,n_2|\left(\widetilde{V}^{\text{1loop}}_{N^2LO,\,E}+\widetilde{V}^{\text{cont.}}\right)|n_1,n_2\rangle \,,
  \label{n2loh1ep}
\end{eqnarray}
where $|n_1\rangle$ and $|n_2\rangle$ are the hydrogen atom states. The LO term is $\widetilde{W}^{(0)}$. The suppressions of Eqs.~\eqref{treeh1bp}, \eqref{nloh1ep} and \eqref{n2loh1ep} relative to Eq.~\eqref{loh1ep} are $\sqrt{\alpha}$, $\alpha$ and $\alpha\sqrt{\alpha}$ respectively, as indicated by the superindices.

The first two terms of Eq.~\eqref{n2loh1ep} carry divergent pieces. These can be recast into local terms by using the results of Appendix~\ref{sumst} and after $\MS$ subtraction the residual scale dependence cancels against the one
of $d_s$ in $\widetilde{V}^{\text{cont.}}$ (see Eq.~\eqref{dsdv}):
\begin{equation}
  \widetilde{W}^{(d+e)}_{E} + \langle n_1,n_2| \widetilde{V}^{\text{1loop}}_{N^2LO,\,E}|n_1,n_2\rangle  + \langle n_1,n_2| \widetilde{V}^{\text{cont.}}|n_1,n_2\rangle\bigg|_{\log \nu}
  = \frac{8\alpha^2}{3m^2}\log \nu - \frac{14\alpha^2}{3m^2}\log \nu + \frac{2\alpha^2}{m^2}\log \nu  = 0 \,.
\end{equation}
The one-loop contributions including magnetic dipole vertices are strongly suppressed. The first one, involving two electric-magnetic dipole potentials, is $\alpha^3$ suppressed with respect to the LO term $W^{(0)}$. The expressions for the analogous diagrams of Fig.~\ref{h1m2} with magnetic dipole interactions are given in Appendix~\ref{loodi}.

Using the Fourier transforms of Appendix~\ref{foutra}, the van der Waals potential can be written in position space. The LO van der Waals potential, given in Eq.~\eqref{loh1ep}, corresponds to the exchange of two electric dipole potentials and has an $R^{-6}$ dependence in position space 
\begin{equation}
  W^{(0)} = \frac{3}{8\pi^2R^6}\sideset{}{'}\sum_{m_1,\,m_2}\frac{p_E(n_1,m_1)p_E(n_2,m_2)}{\D E_{n_1m_1}+\D E_{n_2m_2}}\,.
  \label{idk1}
\end{equation}
Comparing with the London potential \eqref{london} we obtain
\begin{equation}
C_6 = - \frac{3}{8\pi^2}\sideset{}{'}\sum_{m_1,\,m_2}\frac{p_E(n_1,m_1)p_E(n_2,m_2)}{\D E_{n_1m_1}+\D E_{n_2m_2}}\,.
\end{equation}
If the hydrogen atoms are in the ground state the London potential \eqref{idk1} is attractive, and a numerical evaluation that includes discrete and continuum intermediate states gives $C_6=1.73123 \dots \, 10^{-3}$ keV$^{-5}$~\cite{Sucher:1968}. Moreover, an approximation that holds for the ground state $n_1=n_2=1$,
\begin{equation}
\frac{3}{8}E_1\ge\frac{\D E_{1m_1}\D E_{1m_2}}{\D E_{1m_1}+\D E_{1m_2}}\ge \frac{1}{2}E_1\,,
\end{equation}
where $E_1$ is the ground state energy, allows to write Eq.~\eqref{idk1} in the traditional form obtained by London~\cite{london} using second-order time-independent perturbation theory:
\begin{equation}
W^{(0)} \approx -\frac{3}{2}\frac{w_0\,\alpha^2_{1}}{R^6}\,,
\end{equation}
where $w_0 \approx -E_1/2$ and $\alpha_{1}$ is the polarizability of the ground-state hydrogen atom.

Following the counting, after the LO London potential the most important interaction is given by the magnetic dipole potential $W^{(1/2)}$: 
\begin{equation}
  W^{(1/2)} = \frac{\alpha}{m^2}
  \left[\frac{2\pi}{3}\delta^{(3)}\left(\bm{R}\right)\langle n_1| \bm{\mu} |n_1\rangle\cdot\langle n_2| \bm{\mu} |n_2\rangle
  + \frac{3}{4R^3}\hat{\bm{R}}\cdot\langle n_1| \bm{\mu} |n_1\rangle \, \hat{\bm{R}}\cdot\langle n_2| \bm{\mu} |n_2\rangle\right].
  \label{mdpcs}
\end{equation}
This term does not appear in Ref.~\cite{Feinberg:1989ps}, since only spinless particles were considered there. The magnetic dipole potential can be attractive or repulsive depending on the orientation of the angular momenta of the atoms. 

The first correction to the London potential that does not depend on the intrinsic magnetic dipole moments of the atoms is given by $W^{(1)}$, which reads
\begin{equation}
  W^{(1)} = - \frac{1}{8\pi^2R^4}\sideset{}{'}\sum_{m_1,m_2}p_E(n_1,m_1)p_E(n_2,m_2)\frac{\D E_{n_1m_1}\D E_{n_2m_2}}{\D E_{n_1m_1}+\D E_{n_2m_2}}\,.
  \label{idk2}
\end{equation}
$W^{(1)}$ is also attractive for atoms in the ground state. This subleading term for $S$-wave states was derived by Hirschfelder and Meath~\cite{hirsch} and also by Feinberg, Sucher and Au~\cite{Feinberg:1989ps} using dispersion theory. However, in those previous works the potentials were presented depending on integrals over the Compton scattering form factors of a neutral spinless particle. These form factors can be obtained for $S$-wave hydrogen atoms by adding to the diagram on the left-hand side of Fig.~\ref{h2tree} (with incoming energy $E_n+\omega$) the equivalent one with the photon lines crossed:
\begin{equation}
  F_E(\omega) = \sideset{}{'}\sum_m p_E(n,m)\frac{2\D E_{nm} }{\D E^2_{nm}-\omega^2}\,,\qquad
  F_M(\omega) = \sideset{}{'}\sum_m p_B(n,m)\frac{2\D E_{nm} }{\D E^2_{nm}-\omega^2}\,.
  \label{fdpol}
\end{equation}
Using the form factors of Eq.~\eqref{fdpol} in the formulas of Ref.~\cite{Feinberg:1989ps} for the short-distance potentials we obtain the leading and subleading terms of the London potential of Eqs.~\eqref{loh1ep} and \eqref{nloh1ep} (cf. Appendix~\ref{loodi} for the formulas with magnetic dipoles). The form factors of Eq.~\eqref{fdpol} can be interpreted as a frequency-dependent version of the polarizabilities of Eq.~\eqref{pol0}.

The remaining contribution from $W^{(3/2)}$ contains both a $R^{-3}$ part and a Dirac-delta potential:
\begin{equation}
  W^{(3/2)} =  -\frac{7\alpha^2}{6\pi m^2 R^3}+\dots\,,
  \label{idk3}
\end{equation}
where the dots denote the Dirac-delta piece.

\vspace{\baselineskip}

In this section, we have build the EFTs suited to study the van der Waals interactions in the short-distance regime. This regime is characterized by the time scale between the emission of the two photons, which is of order $1/(m \alpha^2)$, being much larger than the distance between the atoms, $R$. In a first step, we have integrated out modes scaling like $1/R$ and matched pNRQED to pNRQED$^{\prime}$. This leads to the appearance of the well-known electric and magnetic dipole potentials as well as to subleading velocity-dependent potentials. Loop contributions, stemming from two-photon exchanges, contribute to the pNRQED$^{\prime}$ two-atom potential at N$^2$LO. In a second step, we have integrated out modes scaling like $m \alpha^2$ and matched pNRQED$^{\prime}$ to WEFT obtaining the van der Waals potentials. The exchange of two electric-dipole potentials in Eq.~\eqref{idk1} corresponds to the London potential. The N$^2$LO van der Waals potential in Eq.~\eqref{idk2} is obtained by considering the exchange of one leading and one subleading dipole potential. The N$^2$LO pNRQED$^{\prime}$ two-atom potential trivially matches into the N$^3$LO van der Waals potential, and turns out to contain a previously unknown $R^{-3}$ term shown in Eq.~\eqref{idk3}, and a local term. Further subleading matching contributions produce only local terms that are however crucial for renormalization.

\section{Long-range Van der Waals interactions}
\label{lrvdwi}
\begin{figure}[ht]
\begin{tabular}{cc}
\includegraphics[width=.3\textwidth]{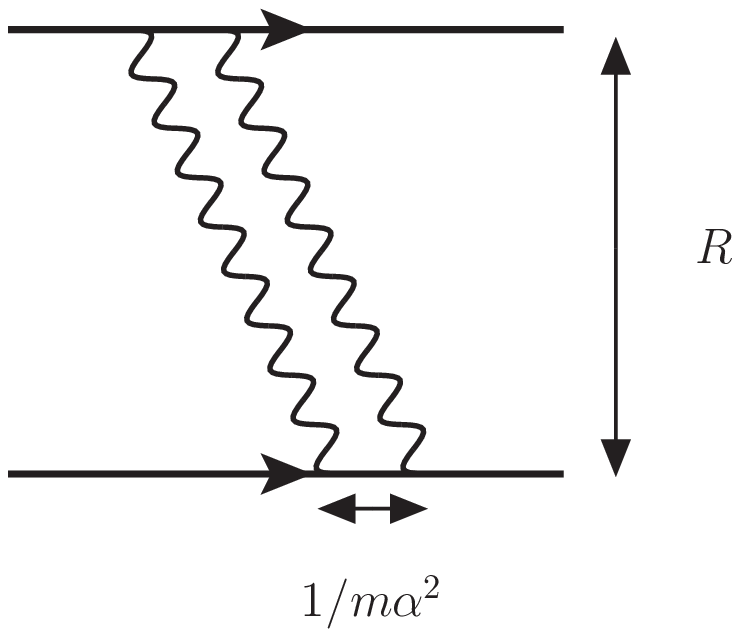}  &  \hspace{2cm}\raisebox{0.7cm}{\includegraphics[width=.25\textwidth]{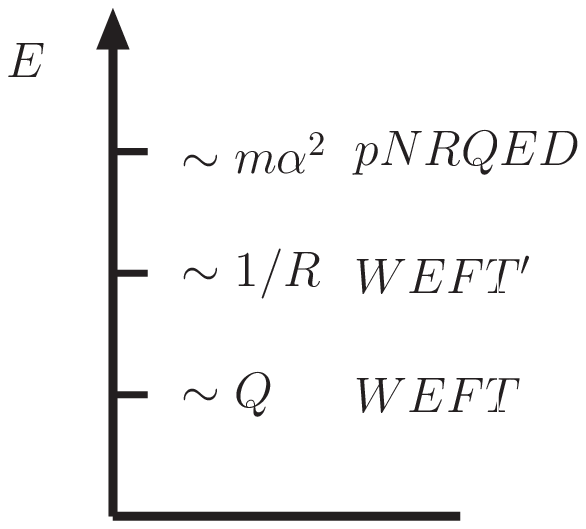}}
\end{tabular}
\caption{Left panel: Sketch of the physical picture of the van der Waals interactions in the long-range regime. The distance between the two atoms is much larger than the typical time interval in which the photons are exchanged. Right panel: Hierarchy of scales and the corresponding EFTs in the long-range regime.}
\label{h2sk}
\end{figure}

In this section, a different physical setting is explored. We consider the case of the long-distance van der Waals interactions. We have sketched the physical picture in the left-hand panel of Fig.~\ref{h2sk}. In this regime the distance between the atoms is much larger than the time scale between the emission of the two photons, which is of the order of the inverse of the ultrasoft scale: $R \gg 1/(m \alpha^2)$. Photons exchanged between the atoms carry a typical momentum, $\bm{k}$, that is of the order of $m \alpha^2$ or the inverse of the distance between the atoms or smaller. Again, the energy scale of the atoms, $Q \sim \bm{k}^2/M$, is much smaller than any other scale due to the strong suppression in $M$. In the right-hand panel of Fig.~\ref{h2sk} we show the hierarchy of scales in the long-distance regime together with the suitable EFT at each scale. The results in this section are valid for arbitrarily long distances, since the hierarchy between the scales in the right-hand panel of Fig.~\ref{h2sk} does not change as the distance $R$ increases.

In the following sections, we will provide the details of the matching between the hierarchy of EFTs in the right-hand panel of Fig.~\ref{h2sk}. In Sec.~\ref{h2vdwp} we integrate out ultrasoft photons carrying energy and momentum of order $m\alpha^2$ and virtual atomic states, whose energy is also of order $m\alpha^2$, and obtain WEFT$^{\prime}$. In the one-atom sector WEFT$^{\prime}$ is equivalent to WEFT (see Sec.~\ref{vdweftsec}). In the two-atom sector WEFT$^{\prime}$ differs from WEFT in that photons with momenta of order $1/R$ are still dynamical. The van der Waals potential between hydrogen atoms in $S$-wave states is obtained in Sec.~\ref{h2mweft} by integrating out photons with momenta of order $1/R$ and matching WEFT$^{\prime}$ to WEFT.

In order to make the counting homogeneous, it is convenient to assign a specific size to $1/R$ in terms of $m$ and $\alpha$. A natural choice, given the scale hierarchy in the long-distance regime, is to take $1/R \sim m \alpha^2 \sqrt{\alpha}$. From this assignment it follows that $ (m\alpha^2\, R) \sim 1/\sqrt{\alpha}$. In the long-distance regime, we aim at computing the non-local van der Waals interactions up to order $m \alpha^{11}\sqrt{\alpha}$ in coordinate space.

\subsection{Matching WEFT\texorpdfstring{$^{\prime}$}{'}}
\label{h2vdwp}
To aid in the computation of the van der Waals potential we introduce WEFT$^{\prime}$, an EFT for momenta much smaller than the typical binding energies, but of the same order as the inverse distance between hydrogen atoms. WEFT$^{\prime}$ follows from pNRQED by integrating out ultrasoft photons carrying energy and momentum of order $m\alpha^2$ and virtual atomic states, whose energy is also of order $m\alpha^2$. The Lagrangian for WEFT$^{\prime}$ in the one-atom sector is the same as for WEFT, and is given in Eq.~\eqref{aeft1sl}. We will now compute the two-atom sector of WEFT$^{\prime}$.

Since we are integrating out photons and virtual atomic states carrying an energy of order $m\alpha^2$, and since the energy scale $m\alpha^2$ is generated only in loops if the initial- and final-state atoms have the same quantum numbers, the only tree-level contributions to consider are potentials taken over from pNRQED to WEFT$^{\prime}$. The leading potential from pNRQED is the contact interaction of Eq.~\eqref{2sci}, which gives 
\begin{equation}
  (\widetilde{W}')^{\text{cont.}} = - \frac{d_s}{m^2} - \frac{4d_v}{m^2} \langle n_1| \bm{S} |n_1\rangle \cdot \langle n_2| \bm{S} |n_2\rangle \,.
\label{contWEFTprime}
\end{equation}

\begin{figure}[ht]
\centering{\includegraphics[height=2.3cm]{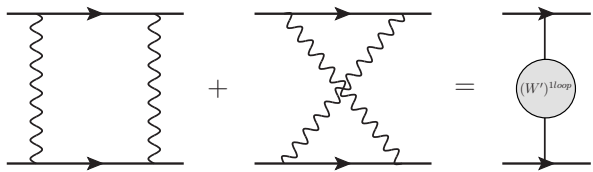}}
\caption{One-loop matching of the two hydrogen atom potentials of WEFT$^{\prime}$. The pNRQED diagrams are on the left-hand side and the WEFT$^{\prime}$ one is on the right-hand side .}
\label{h21}
\end{figure}

The dominant one-loop contributions to the two-atom potential of WEFT$^{\prime}$ are given by the two-photon exchange diagrams in the left-hand side of Fig.~\ref{h21}. The two photons are ultrasoft, which means that they carry a momentum $q^\mu$ that scales like $q^0\sim |\bm{q}|\sim m\alpha^2$. The LO contribution, involving four electric-dipole vertices is of order $\alpha^2/m^2$ in momentum space ($m\alpha^9\sqrt{\alpha}$ in coordinate space according to the counting $1/R \sim m \alpha^2 \sqrt{\alpha}$). Subsequent contributions are suppressed by powers of $1/(m\alpha^2 \,R)^2 \sim \alpha$. Furthermore, replacing an electric dipole coupling by a magnetic one adds at least an extra $\alpha$ suppression.

As we will see in the next section, the Casimir-Polder potential is generated by the one-loop diagram with two-photon exchange through the electric-polarizability seagull vertices of WEFT$^{\prime}$ (fourth diagram in Fig.~\ref{h22}), and it is $\alpha^2$ suppressed with respect to the LO contribution. To match that precision we have to compute the one-loop diagrams of Fig.~\ref{h21} up to N$^2$LO. The different contributions to the WEFT$^{\prime}$ potential for $S$-wave states read in momentum space
\begin{eqnarray}
  (\widetilde{W}')^{\text{1loop}}_{LO,\,E}&=&-(d^2-5d+6)\sideset{}{'}\sum_{m_1,m_2}\frac{p_E(n_1,m1)p_E(n_2,m_2)}{\D E^2_{n_1m_1}-\D E^2_{n_2m_2}}
  \nonumber\\
  && \hspace{3cm} \times  \D E_{n_1m_1}\D E_{n_2m_2} \left[\D E_{n_1m_1} J(\D E_{n_1m_1}) -\D E_{n_2m_2} J(\D E_{n_2m_2})\right],
  \label{loh2}\\
  (\widetilde{W}')^{\text{1loop}}_{NLO,\,E}&=&-(d-2)(d^2-8d+27)\frac{\bm{k}^2}{12}\sideset{}{'}\sum_{m_1,m_2}\frac{p_E(n_1,m_1)p_E(n_2,m_2)}{\D E^2_{n_1m_1}-\D E^2_{n_2m_2}}
  \nonumber\\
  && \hspace{3cm} \times \left[\D E_{n_1m_1} J(\D E_{n_2m_2}) -\D E_{n_2m_2} J(\D E_{n_1m_1})\right],
  \label{nloh2}\\
  (\widetilde{W}')^{\text{1loop}}_{N^2LO,\,E}&=&(d-3)(d-2)(d^2-12d+55)\frac{\bm{k}^4}{240}\sideset{}{'}\sum_{m_1,m_2}\frac{p_E(n_1,m_1)p_E(n_2,m_2)}{\D E^2_{n_1m_1}\D E^2_{n_2m_2}
    \left[\D E^2_{n_1m_1}-\D E^2_{n_2m_2}\right]}
  \nonumber \\
  && \hspace{3cm} \times \left[\D E^3_{n_1m_1} J(\D E_{n_2m_2})-\D E^3_{n_2m_2} J(\D E_{n_1m_1})\right],
  \label{n2loh2} \\
  (\widetilde{W}')^{\text{1loop}}_{LO,\,M}&=&-(d-2)\sideset{}{'}\sum_{m_1,m_2}\frac{p_E(n_1,m_1)p_B(n_2,m_2)_{ii}+p_B(n_1,m_1)_{ii} p_E(n_2,m_2)}{\D E^2_{n_1m_1}-\D E^2_{n_2m_2}}
  \nonumber \\
  && \hspace{3cm} \times \D E_{n_1m_1}\D E_{n_2m_2}\left[\D E_{n_1m_1} J(\D E_{n_1m_1})-\D E_{n_2m_2} J(\D E_{n_2m_2})\right],
  \label{lobeh2}
\end{eqnarray}
where $J$ is a loop integral whose explicit expression can be found in Appendix~\ref{looi}, $d$ is the space-time dimension and summation over the index $i$ is understood. We have used the subscripts $E$ and $M$ to indicate that the contribution is generated in pNRQED by four electric dipole couplings, and two electric and two magnetic dipoles, respectively.

\subsection{Matching WEFT}
\label{h2mweft}
\begin{figure}[ht]
\centering{\includegraphics[height=2.3cm]{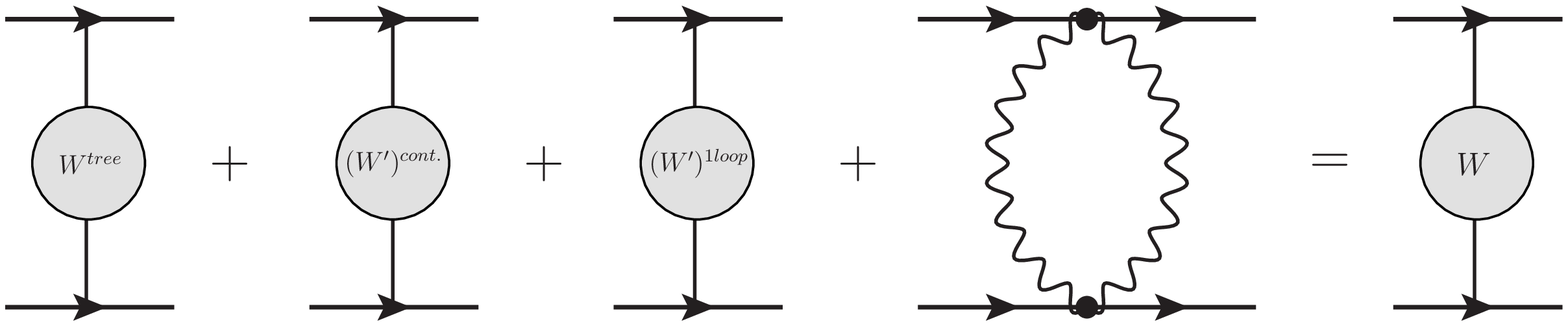}}
\caption{Matching of the van der Waals potential between two hydrogen atoms in the long-range regime. The WEFT$^{\prime}$ diagrams are on the left-hand side and the WEFT one is on the right-hand side.}
\label{h22}
\end{figure}

The matching of WEFT$^{\prime}$ to WEFT consists in integrating out photons with momenta scaling like $1/R$. This is shown diagrammatically in Fig.~\ref{h22}.

The first contribution to the two-atom potential of WEFT is given by the one-photon exchange diagram when the momentum transfer is of order $1/R$. Since we are interested in the case when initial- and final-state atoms have the same quantum numbers, the energy transferred by the photon in the tree-level diagram is zero. Furthermore the electric dipole vertex vanishes when evaluated between initial and final states that are equal. Hence the only contribution comes from the two magnetic dipole potential: 
\begin{equation}
  \widetilde{W}^{\text{tree}} =
  \frac{e^2}{4m^2}\left(\langle n_1| \bm{\mu} |n_1\rangle\cdot\langle n_2| \bm{\mu} |n_2\rangle
  - \frac{\langle n_1| \bm{\mu} |n_1\rangle\cdot\bm{k}\,\langle n_2| \bm{\mu} |n_2\rangle\cdot\bm{k}}{\bm{k}^2}\right).
\label{WtreeBB}
\end{equation}
The second and third contributions take over the potentials \eqref{contWEFTprime}-\eqref{lobeh2} of WEFT$^{\prime}$. The fourth contribution is a one-loop diagram in WEFT$^\prime$ made of the seagull vertices defined on the right-hand side of Fig.~\ref{h2tree}. Further higher-order contact terms like radiative corrections to the matching coefficients $d_s$ and $d_v$, higher-order terms in the multipole expansion of the four-electron operators of NRQED or four-electron operators of dimension eight have not been displayed.

The different contributions to the WEFT potential for $S$-wave states read in momentum space 
\begin{eqnarray}
  \widetilde{W}^{(-1)}&=& \widetilde{W}^{\text{tree}}\,,
  \\
  \widetilde{W}^{(0)}&=&(\widetilde{W}')^{\text{1loop}}_{LO,\,E} + (\widetilde{W}')^{\text{cont.}}
  \label{wh2loe}\,,
  \\
  \widetilde{W}^{(1)}&=&(\widetilde{W}')^{\text{1loop}}_{NLO,\,E} + \dots \,,
  \label{wh2nloe}
  \\
  \widetilde{W}^{(2)}&=&(\widetilde{W}')^{\text{1loop}}_{N^2LO,\,E} + \widetilde{W}_E^{\text{seg}} + (\widetilde{W}')^{\text{1loop}}_{LO,\,M} + \dots \,,
  \label{wh2n2loe}
\end{eqnarray}
where the dots stand for the higher-order contact interactions that have not been computed here. The superindex in brackets indicates the suppression in powers of $\alpha$ with respect to Eq.~\eqref{wh2loe}, which is of order $\alpha^2/m^2$ in momentum space and of order $m \alpha^9\sqrt{\alpha}$ in coordinate space.

The term $\widetilde{W}_E^{\text{seg}}$ is the contribution from the fourth diagram of Fig.~\ref{h22}. The photon momenta and energies scale like~$1/R$. In dimensional regularization $\widetilde{W}_E^{\text{seg}}$ reads in momentum space
\begin{equation}
\widetilde{W}_E^{\text{seg}}= -\frac{(d-2)(4d+7)\pi^2}{8(d-1)(d+1)}\alpha_{n_1}\alpha_{n_2}\bm{k}^4 A_{3/2}(\bm{k}^2),
\label{h2seag1}
\end{equation}
where $\alpha_{n}$ is the electric polarizability of the hydrogen atom as defined in Sec.~\ref{vdweftsec}, $A_{3/2}$ can be found in Appendix~\ref{looi} and $d$ is the space-time dimension.

Ultraviolet divergences are present in Eqs.~\eqref{wh2loe}-\eqref{wh2n2loe}. In the case of the seagull diagram, the divergence and scale dependence in $\widetilde{W}_E^{\text{seg}}$ cancels with the corresponding ones in $(\widetilde{W}')^{\text{1loop}}_{N^2LO,\,E}$:
\begin{equation}
\begin{split}
  \widetilde{W}_E^{\text{seg}} + (\widetilde{W}')^{\text{1loop}}_{N^2LO,\,E}\bigg|_{1/(4-d),\,\log \nu}
  = &- \frac{46}{240} \bm{k}^4 \alpha_{n_1}\alpha_{n_2} \left( \frac{1}{4-d} + \log \nu\right)
  \\
  &+ \frac{46}{240} \bm{k}^4  \frac{1}{4\pi^2} \sideset{}{'}\sum_{m_1,m_2}\frac{p_E(n_1,m_1)}{\D E_{n_1m_1}}\frac{p_E(n_2,m_2)}{\D E_{n_2m_2}} \left( \frac{1}{4-d} + \log \nu\right)  = 0 \,.
\end{split}
\end{equation}
The divergence in $(\widetilde{W}')^{\text{1loop}}_{LO,\,E}$ can be recast as a local term when summing over the intermediate states (see Appendix~\ref{sumst}) and, once $\MS$ renormalized, its scale dependence cancels against that one of $(\widetilde{W}')^{\text{cont.}}$:
\begin{equation}
  (\widetilde{W}')^{\text{1loop}}_{LO,\,E} + (\widetilde{W}')^{\text{cont.}}\bigg|_{\log \nu}
  = - \frac{2\alpha^2}{m^2}\log \nu + \frac{2\alpha^2}{m^2}\log \nu  = 0 \,.
\end{equation}
The divergences in $(\widetilde{W}')^{\text{1loop}}_{NLO,\,E}$ carried by the $J$ loop integrals cancel each other making $(\widetilde{W}')^{\text{1loop}}_{NLO,\,E}$ finite. Finally, divergences in $(\widetilde{W}')^{\text{1loop}}_{LO,\,M}$, are at least of order $\alpha^6/m^2$ and hence beyond our accuracy.

The position-space representation of the potentials can be obtained using the results of Appendix~\ref{foutra}. All contributions proportional to positive even powers of the transfer momentum are proportional to Dirac-delta potentials except for the ones that contain a $\log \bm{k}^2$. Therefore, for long-distance van der Waals interactions the only nonlocal term are $W_E^{\text{seg}}$
and the magnetic dipole potential, $W^{(-1)}$, given in coordinate space in Eq.~\eqref{mdpcs}.

The part of $W_E^{\text{seg}}$ containing $\log \bm{k}^2$ is proportional to $R^{-7}$, whereas the part containing the finite pieces of the loop integral is proportional to a Dirac delta in position space. The former corresponds to the van der Waals potential derived by Casimir and Polder by using two-photon exchange and fourth-order noncovariant perturbation theory~\cite{caspol}:
\begin{equation}
  W_E^{\text{\text{seg}}}  = -\frac{23}{4\pi R^7}\alpha_{n_1}\alpha_{n_2}+\dots\,.
\label{holseexp}
\end{equation}
Comparing with the Casimir-Polder potential \eqref{CP} we obtain 
\begin{equation}
C_7 = \frac{23}{4\pi}\alpha_{n_1}\alpha_{n_2}\,.
\end{equation}

A derivation of the Casimir-Polder potential in Eq.~\eqref{holseexp} using dispersive methods was given by Feinberg and Sucher in Refs.~\cite{Feinberg:1970zz,Feinberg:1989ps}. Feinberg and Sucher also provided the long-range potentials due to magnetic polarizabilities and mixed interactions between electric and magnetic polarizabilities. These can be recovered respectively from our results for $W_B^{\text{seg}}$ and $W_M^{\text{seg}}$ in Appendix~\ref{loodi}. Assuming scalar magnetic polarizabilities, $\beta_n^{ij}=\beta_n \delta^{ij}$, we can write $ \displaystyle W_B^{\text{\text{seg}}} =-\frac{23}{4\pi R^7}\beta_{n_1}\beta_{n_2}+\dots$ and $ \displaystyle W_M^{\text{\text{seg}}} = \frac{7}{4\pi R^7}\left(\alpha_{n_1}\beta_{n_2}+\beta_{n_1}\alpha_{n_2}\right)+\dots\,$. According to our counting, these two cases are suppressed by a factor $\alpha^4$ and $\alpha^2$ respectively compared to $W_E^{\text{\text{seg}}}$. We note, however, that the magnetic polarizability of a hydrogen atom cannot be a scalar since hydrogen possesses a permanent magnetic dipole. The above results for $W_E^{\text{\text{seg}}}$, $W_B^{\text{\text{seg}}}$ and $W_M^{\text{\text{seg}}}$ were also obtained by Holstein~\cite{Holstein:2008fs} using a phenomenological Hamiltonian for the Compton scattering of neutral scalar particles constrained by gauge symmetry, invariance under parity and time reversal.

\vspace{\baselineskip}

We have considered the van der Waals interactions in the long-distance regime. In this case, the distance between the atoms, $R$, is much larger than the time scale, $1/(m \alpha^2)$, between the emission of the two photons. First, we have integrated out ultrasoft photons carrying energy and momentum of order $m\alpha^2$ and matched pNRQED to WEFT$^{\prime}$. The matching produces the polarizability operators in the one-atom sector, and several new local terms in the two-atom sector. In a second step, modes scaling like $1/R$ have been integrated out and the van der Waals potential has been generated as a matching coefficient of WEFT. The two-photon exchange induced by the polarizability operators produces the Casimir-Polder potential~\eqref{holseexp}. The new local terms cancel all ultraviolet divergences in the Casimir-Polder diagram. These results are valid for arbitrarily long distances, since the hierarchy between the scales does not change as the distance $R$ increases.

\section{Intermediate-range van der Waals interactions}
\label{h3}
\begin{figure}[ht]
\begin{tabular}{cc}
\includegraphics[width=.3\textwidth]{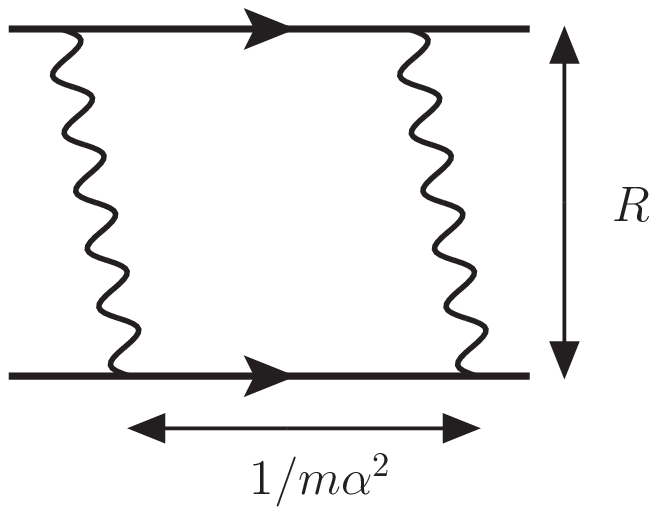}  & \hspace{2cm}\raisebox{0.7cm}{\includegraphics[width=.3\textwidth]{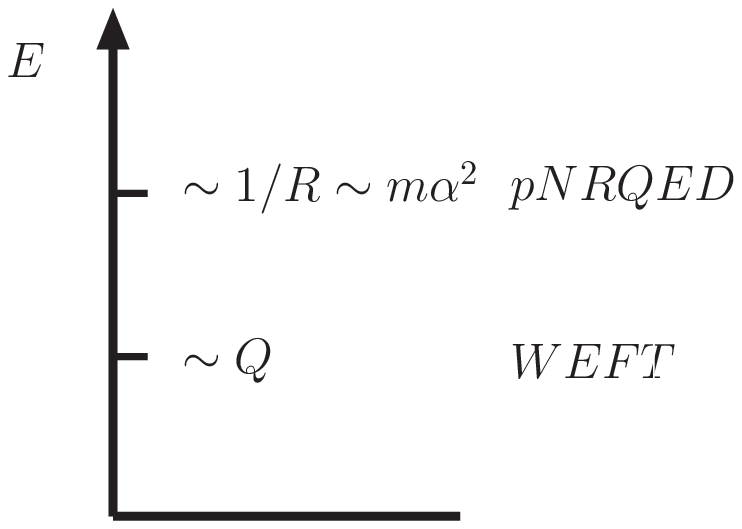}}
\end{tabular}
\caption{Left panel: Sketch of the physical picture of the van der Waals interactions in the intermediate-range regime. The distance between the two atoms is of the same size as the intrinsic time scale of the hydrogen atom. Right panel: Hierarchy of scales and the corresponding EFTs in the intermediate-distance regime.}
\label{h3sk}
\end{figure}

The last possible physical situation to consider is when the range of the van der Waals interactions, $R$, is of the same order as the intrinsic time scale of the hydrogen atom, $1/(m\alpha^2)$. A visual representation of this case is sketched in the left panel of Fig.~\ref{h3sk}. In this regime the distance between the atoms is of the same order as the time scale between the emission of the two photons, which is of the order of the inverse of the ultrasoft scale: $R \sim 1/(m \alpha^2)$. The hierarchy of the two energy scales in the intermediate-distance regime is plotted in the right-hand panel of Fig.~\ref{h3sk}. In this section we obtain the van der Waals potential by integrating out at the same time photons and virtual atomic states with momenta and energies of order $1/R$ and $m\alpha^2$, and matching pNRQED directly to WEFT.

\begin{figure}[ht]
\centering{\includegraphics[height=2.3cm]{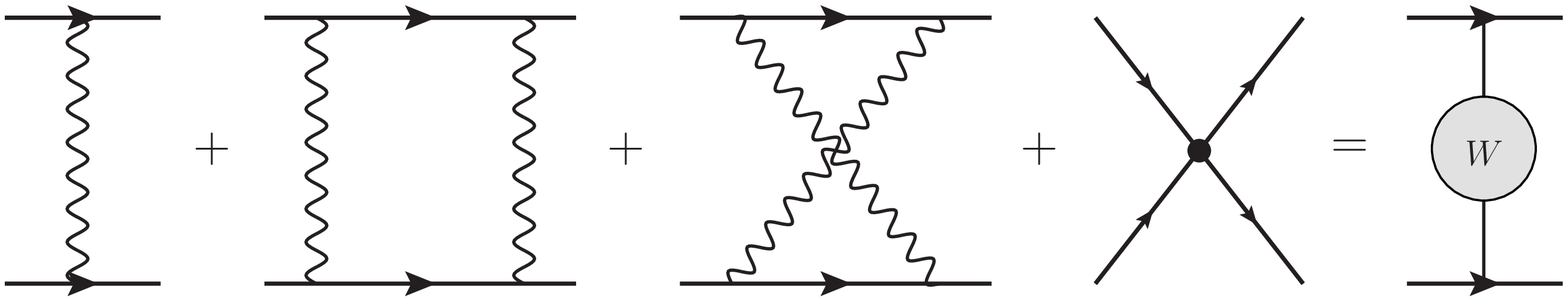}}
\caption{Matching of the van der Waals potential between two hydrogen atoms in the intermediate-range regime. The pNRQED diagrams are on the left-hand side and the WEFT one on the right-hand side.}
\label{h3m}
\end{figure}

The diagrams involved in the matching are shown in Fig.~\ref{h3m}. The dominant contribution for interactions between atoms in $S$-wave states is given by a photon exchange between the permanent magnetic dipoles. The expression is in Eq.~\eqref{WtreeBB}. It is of $\mathcal{O}(\alpha/m^2)$ in momentum space and of $\mathcal{O}(m\alpha^7)$ in coordinate space.

The contribution of the one-loop pNRQED diagrams is
\begin{eqnarray}
\widetilde{W}^{\text{1loop}}_E &=& \sideset{}{'}\sum_{m_1,m_2}p_E(n_1,m_1) p_E(n_2,m_2)
\bigg\{ \frac{1}{2\left(\D E^2_{n_1 m_1}-\D E^2_{n_2 m_2}\right)} \big[\bm{k}^2\left(\D E_{n_2 m_2} J(\D E_{n_1 m_1}) -\D E_{n_1 m_1} J(\D E_{n_2 m_2})\right) 
\nonumber \\
&& \hspace{3cm} 
- \left(\bm{k}^4-4\bm{k}^2\D E^2_{n_1 m_1}+4(d-2)\D E^4_{n_1 m_1}\right)\D E_{n_2 m_2} K(\bm{k}^2,\,\D E_{n_1 m_1})
\nonumber \\
&& \hspace{3cm} 
+ \left(\bm{k}^4 -4\bm{k}^2\D E^2_{n_2 m_2}+4(d-2)\D E^4_{n_2 m_2}\right)\D E_{n_1 m_1} K(\bm{k}^2,\,\D E_{n_2 m_2})\big] 
\nonumber \\
&&\hspace{4.5cm}-\frac{d-2}{2}\D E_{n_1 m_1}\D E_{n_2 m_2} A_{3/2} (\bm{k}^2) \bigg\}.
\label{mrol}
\end{eqnarray}
The explicit definitions of the loop integrals $J$, $K$ and $A_{3/2}$ can be found in Appendix~\ref{looi}. 

The remaining contribution is the contact term
\begin{equation}
  \widetilde{W}^{\text{cont.}} = - \frac{d_s}{m^2}-\frac{4d_v}{m^2} \langle n_1| \bm{S} |n_1\rangle \cdot \langle n_2| \bm{S} |n_2\rangle \,.
\end{equation}
The one-loop electric dipole diagrams and the contact term are of $\mathcal{O}(\alpha^2/m^2)$ in momentum space and of $\mathcal{O}(m\alpha^8)$ in coordinate space. The rest of the terms are suppressed by one power of $\alpha$ for each magnetic dipole replacing an electric dipole. 

In summary, we have that 
\begin{eqnarray}
\widetilde{W}^{(-1)}&=&\widetilde{W}^{\text{tree}}\,,
\\
\widetilde{W}^{(0)}&=&\widetilde{W}^{\text{1loop}}_E+\widetilde{W}^{\text{cont.}}\,, 
\label{h3vpe}
\end{eqnarray}
where the superindex indicates that $\widetilde{W}^{(0)}$ is suppressed by one power of $\alpha$ compared to  $\widetilde{W}^{(-1)}$. The ultraviolet divergence in $\widetilde{W}^{\text{1loop}}$ can be recast as a local term when summing over the intermediate states (see Appendix~\ref{sumst}) and, once $\MS$ renormalized, its scale dependence cancels against that one of $\widetilde{W}^{\text{cont.}}$:
\begin{equation}
  \widetilde{W}^{\text{1loop}}_E + \widetilde{W}^{\text{cont.}}\bigg|_{\log \nu}
  = - \frac{2\alpha^2}{m^2}\log \nu + \frac{2\alpha^2}{m^2}\log \nu  = 0 \,.
\end{equation}

As a cross-check, the expression in Eq.~\eqref{mrol} can be expanded in powers of $\D E_{nm}/|\bm{k}|$, yielding the short-distance van der Waals interactions of Eqs.~\eqref{h1helo}-\eqref{deh1} from Sec.~\ref{srvdwi}. Analogously, expanding in powers of $|\bm{k}|/\D E_{nm}$ results in the long-distance van der Waals expressions of Eqs.~\eqref{loh2}-\eqref{n2loh2} and Eq.~\eqref{h2seag1} from Sec.~\ref{lrvdwi}.

The potential in coordinate space is obtained by Fourier transforming Eq.~\eqref{h3vpe}. However, in the one-loop term $\widetilde{W}^{\text{1loop}}_E$ there are nonanalytic pieces that cannot be transformed using the results of Appendix~\ref{foutra}. The coordinate-space potential associated to these pieces can be obtained by using a dispersive representation of the momentum space potential (see, e.g., Refs.~\cite{Feinberg:1989ps,Brambilla:2015rqa}).

The coordinate space potential is given by 
\begin{equation}
 W(R)= \int \frac{d^3 k}{(2 \pi)^3} \, e^{i \bm{k} \cdot \bm{R}} \, \widetilde{W} (k)\,. 
 \label{ft}
\end{equation}
Since for $\bm{k}^2\rightarrow \infty$ the momentum-space potential $\widetilde{W}^{\text{1loop}}_E$ diverges as $\bm{k}^4$, its corresponding dispersion relation should be twice subtracted. 
The subtraction constants are independent of the momentum and as such correspond to Dirac-delta potentials. The nonlocal part of the potential is given by the following dispersive representation corresponding to the two-photon cut
\begin{equation}
  \widetilde{W}(k)=\frac{2}{\pi}\int^{\infty}_0 d\mu \, \frac{\mu\, \text{Im} \left[ \widetilde{W}(\eta-i\mu) \right]}{\mu^2+k^2} \,,
  \label{esprep}
\end{equation}
where the limit $\eta\rightarrow 0$ is understood. Plugging Eq.~\eqref{esprep} into Eq.~\eqref{ft} and changing the order of the dispersive and Fourier integrals we arrive at
\begin{equation}
  W(R)=\frac{1}{2\pi^2 R}\int^{\infty}_{0}d\mu\,e^{-\mu R}\mu\,\text{Im}\left[\widetilde{W}(\eta-i\mu)\right].
  \label{idk4}
\end{equation}
The imaginary part of $\widetilde{W}^{\text{1loop}}_E$ can be easily obtained after inserting the explicit values of the loop integrals $J$, $K$ and $A_{3/2}$ from Appendix~\ref{looi} into Eq.~\eqref{mrol}. 
This yields
\begin{eqnarray}
\text{Im}\left[\widetilde{W}^{\text{1loop}}_E(\eta-i\mu)\right]&=&
-\sideset{}{'}\sum_{m_1,m_2}\frac{p_E(n_1,m_1) p_E(n_2,m_2)}{16\pi} \bigg\{ 4\D E_{n_1 m1}\D E_{n_2 m_2}
\nonumber \\
& & \hspace{-2cm}
+ \frac{1}{\mu\left(\D E^2_{n_1 m_1}-\D E^2_{n_2 m_2}\right)}
\left[\left(\mu^4  +4\mu^2\D E^2_{n_1 m_1}+8\D E^4_{n_1 m_1}\right)\D E_{n_2 m_2}\arccot\left(\frac{2|\D E_{n_1 m_1}|}{\mu}\right) \right. 
\nonumber \\
& & \hspace{1.8cm}
\left. - \left(\mu^4+4\mu^2\D E^2_{n_2 m_2}+8\D E^4_{n_2 m_2}\right) \D E_{n_1 m_1}\arccot\left(\frac{2|\D E_{n_2 m_2}|}{\mu}\right)\right]\bigg\}.
\label{idk5}
\end{eqnarray}

\begin{figure}[ht]
\begin{tabular}{cc}
\includegraphics[height=6cm]{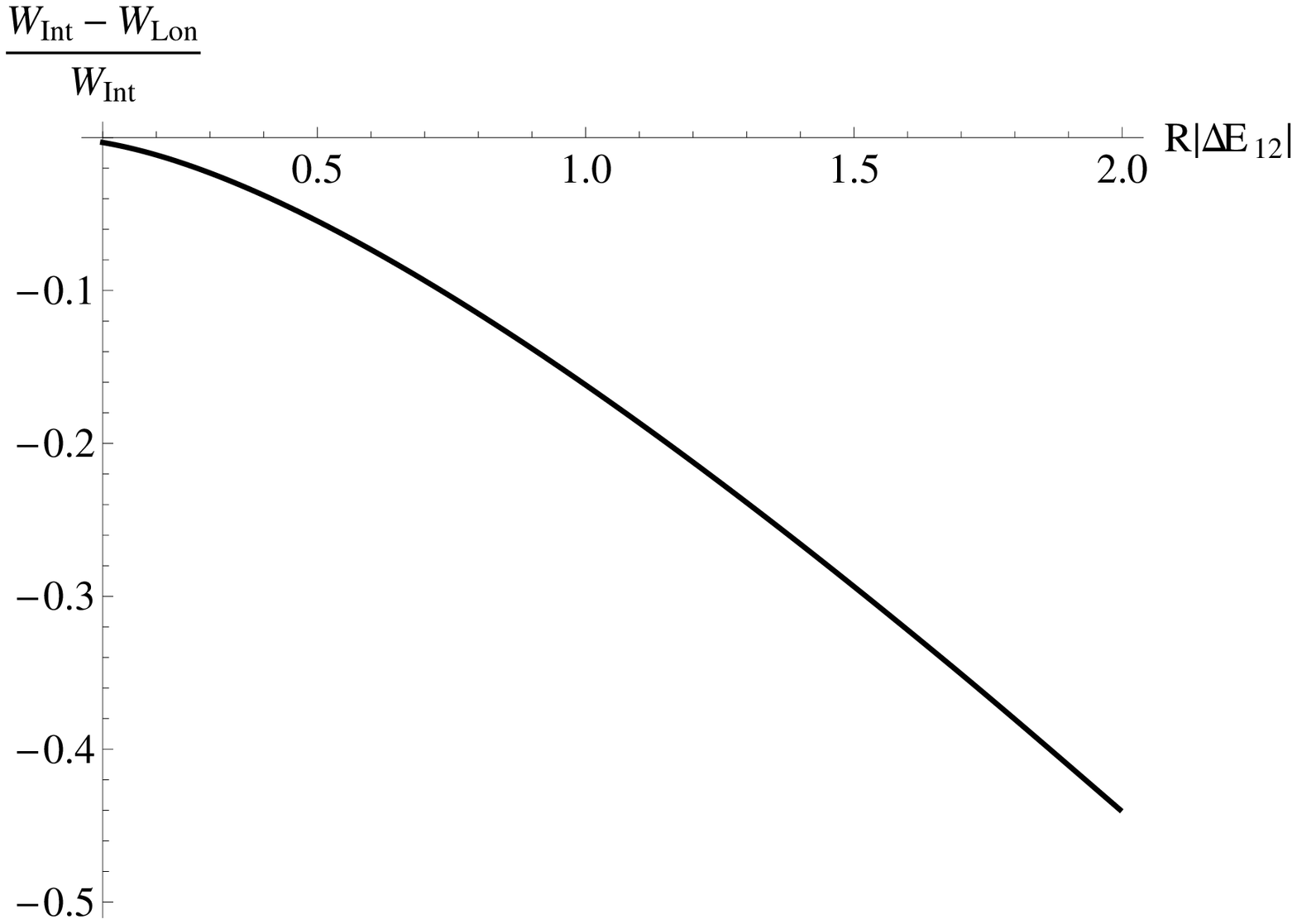} & \includegraphics[height=6cm]{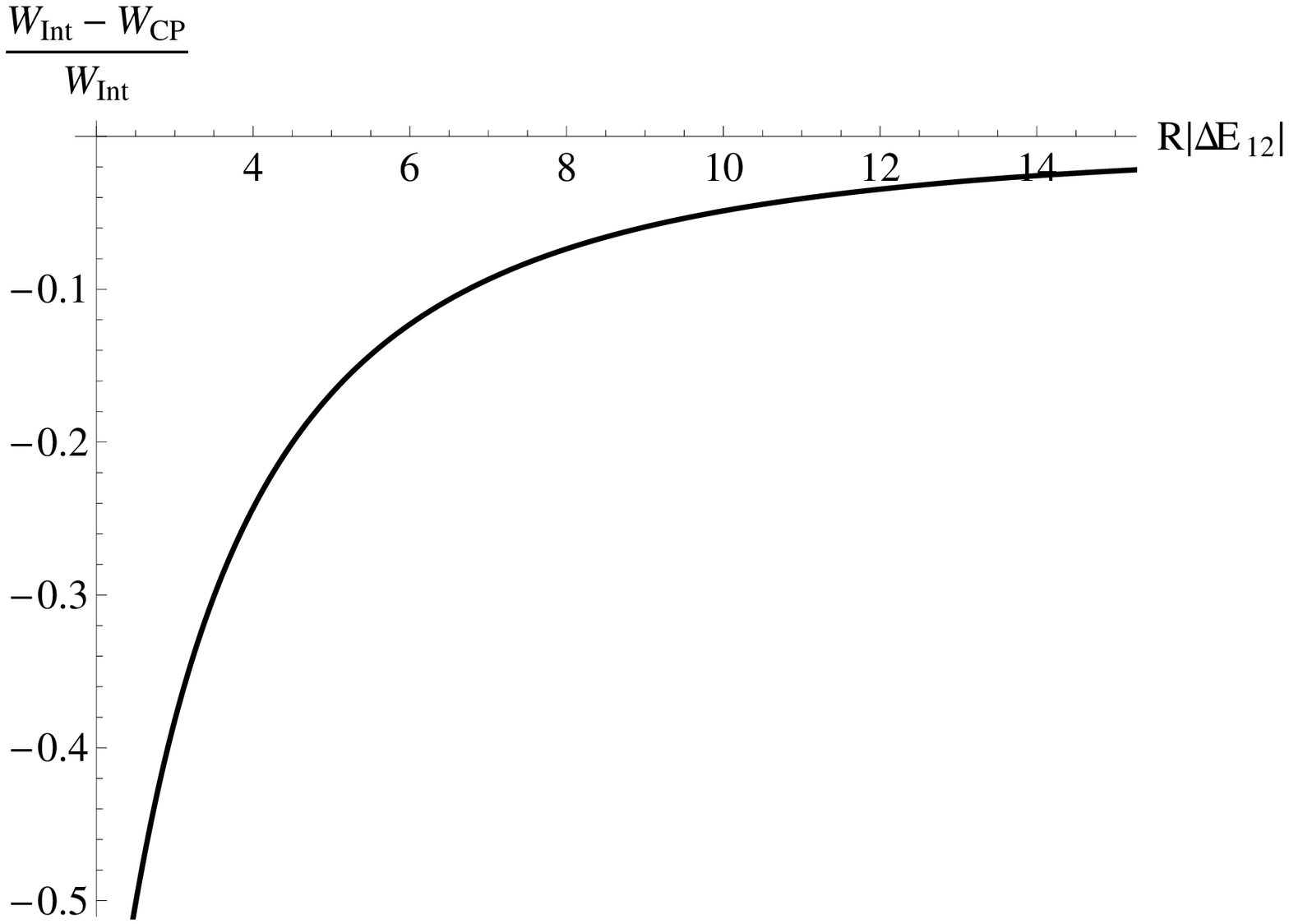} \\
{\it (a)} & {\it (b)} \\
\end{tabular}
\caption{Plots of the relative difference between the intermediate van der Waals potential and the London {\it (a)} and Casimir-Polder {\it (b)} potentials for both atoms in the ground state as a function of the distance $R$ in units of $1/|\D E_{12}|$. $W_{\text{Int}}$ is the intermediate-range potential in Eq.~\eqref{idk4}, $W_{\text{Lon}}$ is the London potential in Eq.~\eqref{idk1} and $W_{\text{CP}}$ is the Casimir-Polder potential in Eq.~\eqref{holseexp}.}
\label{compsl}
\end{figure}

In Fig.~\ref{compsl} we plot the relative difference between the intermediate-range van der Waals potential given in Eqs.~\eqref{idk4} and \eqref{idk5} with the London potential from Eq.~\eqref{idk1} and the Casimir-Polder potential from Eq.~\eqref{holseexp} for both atoms in the ground state. As expected, the London potential and the Casimir-Polder potential are a good approximations in the short and long distances respectively. Due to a conspiracy of numerical factors and cancellations, the convergence towards the London potential is, however, somewhat faster than the one towards the Casimir-Polder potential.

\vspace{\baselineskip}

In summary, in this section we have explored for the first time the intermediate-distance regime $1/R \sim m \alpha^2$ for the van der Waals interaction between two $S$-wave hydrogen atoms. In this regime, pNRQED is directly matched to WEFT, unlike in the short- and long-distance cases where the matching was done in two steps. The van der Waals potential is obtained by integrating out at the same time photons and virtual atomic states with momenta and energies of order $1/R$ and $m\alpha^2$. The ultraviolet divergence in the two-photon exchange diagram that generates the van der Waals potential is removed by contact interactions in the two-atom sector of pNRQED. The coordinate-space representation of the potential in Eq.~\eqref{idk4} is obtained by using a dispersive representation of the momentum space potential in Eq.~\eqref{mrol}. Figure~\ref{compsl} shows the relative difference between the intermediate-range van der Waals potential and the London and the Casimir-Polder potentials 
for both atoms in the ground state as a function of $R$.

\section{Conclusions}
\label{conc}
A hydrogen atom is a nonrelativistic bound state characterized by a hierarchy of well-separated scales. These are the mass of the particle that forms the bound state (hard), the inverse of the Bohr radius, namely the relative momentum (soft), and the typical bound-state energy (ultrasoft). Integrating out the hard scale produces NRQED, and integrating out the soft scale leads to pNRQED, which is the theory best suited to study the bound state.

In systems with two hydrogen atoms two distinct physical regimes exist depending on the distance between the nuclei. If the distance between the nuclei is of the order of the Bohr radius, the system is configured as a diatomic molecule. On the other hand, if this distance is larger than the Bohr radius, the system consists of two atoms interacting through van der Waals interactions. In both cases a new energy scale, different from the intrinsic ones characterizing a single atom, appears. In diatomic molecules, the two-nuclei dynamics takes place at a lower energy scale than the ultrasoft scale in which the electrons bind to the nuclei. This is nothing else than restating the usual assumption that the electron and nuclei dynamics occur at very different time scales, which is at the basis of the Born-Oppenheimer approximation~\cite{BO}. In the van der Waals case, the new scale is the distance between the atoms, which can be larger or smaller than the ultrasoft scale. 

In this work, we have presented a study of the dispersive van der Waals interactions between two hydrogen atoms in the framework of nonrelativistic EFTs of QED. We have focused on $S$-wave states, since these do not have permanent electric multipolar moments and their interaction proceeds through dispersive van der Waals forces in addition to the magnetic-dipole coupling. We have introduced a new EFT, WEFT, to describe the dynamics of the degrees of freedom that live at the low-energy scale where a van der Waals potential is naturally defined. Then the van der Waals potential was obtained by sequentially integrating out the physical scales of the two hydrogen atoms. The EFT setting allows to compute all local terms needed to renormalize the van der Waals interactions, which is the most original result of the present work.

Different hierarchies of scales correspond to different physical scenarios and lead to different results for the dispersive van der Waals potential. We have explored three possible scenarios: short, long and intermediate distances between the atoms. 

In the short-distance regime, $1/R \gg m \alpha^2$, integrating out modes scaling like $1/R$ leads to the well-known electric and magnetic dipole potentials as well as to subleading velocity-dependent potentials. Integrating out the ultrasoft scale leads to the van der Waals potential. The leading contribution of Eq.~\eqref{idk1} stems from the exchange of two electric-dipole potentials and corresponds to the London potential~\cite{london}. Replacing one of the leading dipole potentials with a subleading one, the N$^2$LO term, Eq.~\eqref{idk2}, is obtained. This term is equivalent to the one obtained by Hirschfelder and Meath~\cite{hirsch,Feinberg:1989ps}. In addition we have investigated the N$^3$LO van der Waals potential, previously unknown, which turns out to contain a $R^{-3}$ term, shown in Eq.~\eqref{idk3}, and a local term. 

The long-distance regime corresponds to $1/R \ll m \alpha^2$. Integrating out the ultrasoft scale defines the polarizability operators in the one-atom sector, while in the two-atom sector several local terms are generated. Integrating out the $1/R$ scale, the two-photon exchange induced by the polarizability operators generates the Casimir-Polder potential~\cite{caspol,Feinberg:1970zz,Holstein:2008fs} of Eq.~\eqref{holseexp}. The newly computed local terms turn out to be crucial to cancel all ultraviolet divergences in the Casimir-Polder diagram, which, to our knowledge, has been renormalized in this context here for the first time.

In the last part of the paper, we have explored for the first time the intermediate-distance regime $1/R \sim m \alpha^2$. In this regime the van der Waals potential is obtained by integrating out the ultrasoft and $1/R$ scales simultaneously. We have obtained a coordinate-space representation for the nonlocal part of the van der Waals potential by using a dispersive representation of the momentum space potential. Figure~\ref{compsl} summarizes our findings. It shows the relative difference between the intermediate-range van der Waals potential and the London and the Casimir-Polder potentials for both atoms in the ground state as a function of $R$. The plot shows that the intermediate-range potential is needed to accurately describe (with an accuracy better than about 15\%) the van der Waals interaction in the distance range between $400$ to $2000$ times the Bohr radius. Results for the dispersive van der Waals potentials in the three different regimes have been generalized to states with any angular momentum in Appendix~\ref{loodi}.

We conclude with a possible outlook. The EFT description of the van der Waals interactions between two hydrogen atoms obtained here also offers a framework to appropriately define and systematically calculate also van der Waals interactions for other physical systems starting from the underlying quantum field theory. A prominent case is that of the hadronic van der Waals interactions, whose underlying field theory is QCD. The multigluon interaction is a QCD analogue of the van der Waals force of atomic physics. A color van der Waals force arises in hadron-hadron interactions due to the chromopolarizability of the color-neutral hadrons, similar to the electric polarizability in the case of the hydrogen atom. Contrary to the situation in QED, not much is presently known about color van der  Waals forces: one reason is that they are a long-wavelength feature of QCD and therefore of nonperturbative nature, which makes it difficult to assess them from first principles. The potential relevance of color van der Waals forces for the study of the new hadrons, which may arise as a result of such interaction, demands a better understanding of their properties within QCD.

Quarkonia are hadrons made of a heavy quark and a heavy antiquark. Their hierarchy of energy scales is similar to the one described in this paper for the hydrogen atom, but now the hard, soft and ultrasoft scales are $m_Q$, $m_Q v_Q$ and $m_Q v^2_Q$, with $m_Q$ being the heavy quark mass and $v_Q$ the quark's relative velocity. The velocity $v_Q$ may be identified with the strong coupling constant, $\alpha_{\rm s}$, only if the quarkonium is a Coulombic bound state, which holds for the lowest quarkonium states. Moreover, color provides a richer set of degrees of freedom with respect to QED. In particular, static quark-antiquark pairs may exist at small distances in two possible color configurations. Nevertheless, the hierarchy of EFTs relevant for describing quarkonium-quarkonium interactions is similar to the one discussed in this paper for QED, starting from potential nonrelativistic QCD (pNRQCD)~\cite{Pineda:1997bj,Brambilla:1999xf} to the ultimate van der Waals EFT~\cite{Brambilla:2015rqa}. We have investigated van der Waals interactions for Coulombic quarkonia in~\cite{Brambilla:2015rqa}, whereas van der Waals interactions for nonperturbatively bound quarkonia have been addressed with numerical methods in Ref.~\cite{Bai:2016int}. For long-distance dipole-dipole interactions analytic nonperturbative exact expressions as well as lattice results can be found in Refs.~\cite{Giordano:2009vs,Giordano:2015oza}. EFTs may provide further insights into these systems and link the findings with other processes and systems like quarkonium hadronic transitions, quarkonium-nuclei interactions and exotic multiquark systems.

\acknowledgments

J.T.C. thanks Joan Soto for useful discussions. This work has been supported by the DFG and the NSFC through funds provided to the Sino-German CRC 110 ``Symmetries and the Emergence of Structure in QCD'', and by the DFG cluster of excellence ``Origin and structure of the universe'' (www.universe-cluster.de).

\appendix

\section{Loop Integrals}
\label{looi}
Throughout this work we have used dimensional regularization. We define
\begin{equation}
\mathcal{K}=\bm{q}^2+x(1-x)\bm{k}^2\,.
\end{equation}
The loop integrals that depend only on $\bm{k}^2$ are of the form
\begin{equation}
A_f(\bm{k}^2)=\nu^{4-d}\int^1_0dx\int\frac{d^{(d-1)}q}{(2\pi)^{(d-1)}}\frac{1}{\mathcal{K}^f}\,, 
\end{equation}
where $\nu$ is the renormalization scale. Only $A_{3/2}(\bm{k}^2)$ and $A_{2}(\bm{k}^2)$ appear in our results:
\begin{eqnarray}
  A_{3/2}(\bm{k}^2)&=&\frac{1}{4\pi^2}\left[\lambda+2-\log\left(\frac{\bm{k}^2}{\nu^2}\right)\right], 
  \qquad \hbox{with}\quad \lambda=\frac{2}{4-d}-\gamma_E+\log 4\pi 
  \,,
  \\
  A_{2}(\bm{k}^2)&=&\frac{1}{8|\bm{k}|} \,,
\end{eqnarray}
where here and in the following one-loop results of this appendix [Eqs.~\eqref{jiint} and \eqref{b2int}--\eqref{c2int}] we have neglected terms of $\mathcal{O}(4-d)$ or smaller; $\gamma_E$ is the Euler-Mascheroni constant. The ultraviolet divergence can be renormalized in the $\MS$ scheme by absorbing the pieces proportional to $\lambda$ in the counterterms. In the intermediate calculations other powers in the denominator of $A_f$ may appear. These can be related to $A_{3/2}$ and $A_{2}$ using the following recurrence relation
\begin{equation}
A_{f+1}(\bm{k}^2)=\frac{2(2f-d)}{f\bm{k}^2}A_{f}(\bm{k}^2)\,.
\end{equation}

A loop integral that depends only on $\D E_{nm}$ appears in Secs.~\ref{h1vdeft} and~\ref{h2vdwp}:
\begin{equation}
  J(\D E_{nm})) =
  \nu^{4-d}\int \frac{d^{(d-1)}q}{(2\pi)^{(d-1)}}\frac{1}{2|\bm{q}|(|\bm{q}|-\D E_{nm})}=\frac{\D E_{nm}}{8\pi^2}\left[\lambda+2-2\log 2+2i\pi\theta\left(\D E_{nm}\right)
  -\log\left(\frac{\Delta E^2_n}{\nu^2}\right)\right].
\label{jiint}
\end{equation}

In Sec.~\ref{h3} loop integrals depending simultaneously on $\bm{k}^2$ and $\D E_{nm}$ occur. These can be reduced to the master integrals
\begin{eqnarray}
B_f(\bm{k}^2,\,\D E_{nm})&=&\nu^{4-d}\int^1_0dx\int\frac{d^{(d-1)}q}{(2\pi)^{(d-1)}}\frac{1}{\left(\mathcal{K}-\D E^2_{nm}\right)^f}\,,  
\\
C_f(\bm{k}^2,\,\D E_{nm})&=&\nu^{4-d}\int^1_0dx\int\frac{d^{(d-1)}q}{(2\pi)^{(d-1)}}\frac{1}{\sqrt{\mathcal{K}}\left(\mathcal{K}-\D E^2_{nm}\right)^f}\,,
\end{eqnarray}
which always appear in the combination 
\begin{equation}
  K(\bm{k}^2,\,\D E_{nm}) =
  -\frac{1}{4\D E_{nm}}A_{3/2}(\bm{k}^2)+\frac{1}{2}B_{2}(\bm{k}^2,\,\D E_{nm})+\frac{1}{4\Delta E_{nm}}C_{1}(\bm{k}^2,\,\D E_{nm}) 
  +\frac{\Delta E_{nm}}{2}C_{2}(\bm{k}^2,\,\D E_{nm})\,.
\end{equation}

An explicit analytic result for $B_2$ reads
\begin{equation}
B_2(\bm{k}^2,\,\D E_{nm})=\frac{1}{8\pi |\bm{k}|}\left(\pi+2i\arctanh\sqrt{\frac{4\D E^2_{nm}}{\bm{k}^2}}\right).
\label{b2int}
\end{equation}
This expression is correct in the momentum region $\bm{k}^2 > 4\D E^2_{nm}$. The analytic continuation to the region $\bm{k}^2 < 4\D E^2_{nm}$ is obtained by using the prescription $\D E^2_{nm}\rightarrow\D E^2_{nm}+i\eta$. 

An analytic integration of the Feynman parameters is not possible for $C_1$ and $C_2$. Different expressions for $C_1$ and $C_2$ are possible:
\begin{equation}
C_1(\bm{k}^2,\,\D E_{nm}) = \frac{1}{4\pi^2}\left[\lambda+4-\log \frac{4\D E^2_{nm}}{\nu^2}+i \pi-\int^1_0 dx\frac{1}{\sqrt{1-x}}\sqrt{1-x\frac{4\D E^2_{nm}}{\bm{k}^2}} 
  \arctanh\frac{1}{\sqrt{1-x\frac{4\D E^2_{nm}}{\bm{k}^2}}}\right],
\label{k41f}
\end{equation}
which is valid in the momentum region $\bm{k}^2 > 4\D E^2_{nm}$ and can be analytically continued to the region $\bm{k}^2 < 4\D E^2_{nm}$ by using the prescription $\Delta E^2_{nm}\rightarrow\D E^2_{nm}+i\eta$. The integrand in Eq.~\eqref{k41f} can be expanded for large values of $\bm{k}^2$ but not for small ones. This is because there is always a small enough value of $x$ that makes $x/\bm{k}^2\sim 1$ for any arbitrarily small value of $\bm{k}^2$. To expand for small $\bm{k}^2$ one can use the following expression, which is valid for $\bm{k}^2 < 4\D E^2_{nm}$,
\begin{equation}
C_1(\bm{k}^2,\,\D E_{nm}) = \frac{1}{4\pi^2}\left[\lambda+4-\log\left(\frac{\bm{k}^2}{\nu^2}\right)-\int^1_0 dx\frac{1}{\sqrt{1-x}}\sqrt{1-x\frac{\bm{k}^2}{4\D E^2_{nm}}} 
    \arctanh\frac{1}{\sqrt{1-x\frac{\bm{k}^2}{4\D E^2_{nm}}}}\right].
\end{equation}

The last integral is 
\begin{equation}
  C_2(\bm{k}^2,\,\D E_{nm}) =
  \frac{1}{8\pi^2\D E^2_{nm}}\int^1_0 dx\frac{x\frac{4\D E^2_{nm}}{\bm{k}^2}}{\sqrt{1-x}\sqrt{1-x\frac{4\D E^2_{nm}}{\bm{k}^2}}}\arctanh\frac{1}{\sqrt{1-x\frac{4\D E^2_{nm}}{\bm{k}^2}}}\,.
\label{k31f}
\end{equation}
This expression is correct in the momentum region $\bm{k}^2 > 4\D E^2_{nm}$ and can be expanded for $\bm{k}^2 \gg \D E^2_{nm}$. The analytic continuation to the region $\bm{k}^2 < 4\D E^2_{nm}$ is obtained by using the prescription $\D E^2_{nm}\rightarrow\D E^2_{nm}+i\eta$ and reads  
\begin{equation}
C_2(\bm{k}^2,\,\D E_{nm}) = -\frac{1}{8\pi^2\D E^2_{nm}}
\left[2+\int^1_0 dx\frac{x\frac{\bm{k}^2}{4\D E^2_{nm}}}{\sqrt{1-x}\sqrt{1-x\frac{\bm{k}^2}{4\D E^2_{nm}}}}\arctanh\frac{1}{\sqrt{1-x\frac{\bm{k}^2}{4\D E^2_{nm}}}}\right].
\label{c2int}
\end{equation}
This expression can be expanded for $\bm{k}^2 \ll \D E^2_{nm}$.

\section{Generalized one-loop diagram expressions}
\label{loodi}
In this appendix, we generalize the results of the one-loop contributions with electric dipole interactions to initial and final states of the hydrogen atoms with any value of the angular momentum. Nevertheless, when considering the interaction between two hydrogen atoms in an arbitrary angular momentum state, one should keep in mind that quadrupole and higher multipole moments may not vanish. These multipole couplings can give rise to tree-level interactions that can be as or more important than the one-loop van der Waals potential. For example quadrupole-quadrupole potentials, which appear when both atoms are in a state with $L \ge 2$, are parametrically larger by $\sqrt{\alpha}$ and $\alpha^2$ respectively than the London and Casimir-Polder potentials.

Furthermore, we also provide the expressions for the analogous loop contributions obtained by replacing electric dipoles with magnetic dipoles. From the pNRQED Lagrangian of Eq.~\eqref{pnrqedl} we can see that the magnetic dipole operator is smaller than the electric dipole operator by a factor of order $\alpha$. Since, due to parity, the two couplings on the same atom must either be both magnetic or both electric dipoles this gives two new kinds of loop contributions: one with two electric dipoles and two magnetic dipoles, and one with four magnetic dipoles. In general, all these contributions are much smaller than the ones produced with only electric dipoles. 

Throughout this appendix we use the notation $E$, $B$, $M$ to label contributions from one-loop diagrams with four electric dipoles ($E$), two electric dipoles and two magnetic dipoles ($M$), and four magnetic dipoles ($B$). The explicit expressions for the loop integrals $A_{3/2}$, $A_{2}$, $J$ and $K$ can be found in Appendix~\ref{looi}. Results will be given in $d$ space-time dimensions.

\subsection{Short-range regime}
The one-loop matching contributions from pNRQED to the two-atom pNRQED$^{\prime}$ potential with four electric dipole vertices given in Sec.~\ref{h1mpnrqed} are independent of the initial and final states and thus valid for any angular momentum. We, now, provide the analogous contributions with four magnetic dipole vertices, and with two magnetic dipole vertices on one atom and two electric dipole vertices on the other one. The subscripts 1 and 2 of $\bm{\mu}$ and $\bm{L}$ indicate the atom. In the first case the contributions read 
\begin{eqnarray}
\widetilde{V}^{\text{1loop}}_{LO,B}&=&\frac{\pi^2\alpha^2A_{3/2}(\bm{k}^2)}{16(d-1)m^4}\bm{\mu}_{1\,i} \bm{\mu}_{1\,j} \bm{\mu}_{2\,k} \bm{\mu}_{2\,l}
\nonumber\\
&& 
\times \left[(d-2)(\bm{k}^i\bm{k}^l\delta^{jk}+\bm{k}^j\bm{k}^k\delta^{il}-\bm{k}^i\bm{k}^k\delta^{jl}-\bm{k}^j\bm{k}^l\delta^{ik})
+3\bm{k}^2\left(\delta^{il}\delta^{jk}-\delta^{ik}\delta^{jl}\right)\right],
\\
\widetilde{V}^{\text{1loop}}_{NLO,B}&=&
\frac{\pi^2\alpha^2A_{2}(\bm{k}^2)}{16(d-2)m^4}\left[\left(\bm{\mu}_{1\,i}\dot{\bm{\mu}}_{1\,j}-\dot{\bm{\mu}}_{1\,i}\bm{\mu}_{1\,j}\right)\bm{\mu}_{2\,k}\bm{\mu}_{2\,l}
+\bm{\mu}_{1\,i}\bm{\mu}_{1\,j}\left(\bm{\mu}_{2\,k}\dot{\bm{\mu}}_{2\,l}-\dot{\bm{\mu}}_{2\,k}\bm{\mu}_{2\,l}\right)\right]
\nonumber\\
&& 
\times
\left[\bm{k}^i\bm{k}^k\delta^{jl}+\bm{k}^j\bm{k}^l\delta^{ik}-\bm{k}^i\bm{k}^l\delta^{jk}-\bm{k}^j\bm{k}^k\delta^{il}+2\bm{k}^2\left(\delta^{il}\delta^{jk}-\delta^{ik}\delta^{jl}\right)\right],
\\
\widetilde{V}^{\text{1loop}}_{N^2LO,B}&=&-\frac{\pi^2\alpha^2A_{3/2}(\bm{k}^2)}{12m^4}
\nonumber\\
&& 
\times 
\bigg\{ \bigg[ \dot{\bm{\mu}}_{1\,i}\dot{\bm{\mu}}_{1\,j}\bm{\mu}_{2\,k}\bm{\mu}_{2\,l}+\bm{\mu}_{1\,i}\bm{\mu}_{1\,j}\dot{\bm{\mu}}_{2\,k}\dot{\bm{\mu}}_{2\,l} 
-\frac{1}{4}\left(\bm{\mu}_{1\,i}\dot{\bm{\mu}}_{1\,j}-\dot{\bm{\mu}}_{1\,i}\bm{\mu}_{1\,j}\right)\left(\bm{\mu}_{2\,k}\dot{\bm{\mu}}_{2\,l}-\dot{\bm{\mu}}_{2\,k}\bm{\mu}_{2\,l}\right)\bigg]\lambda_{N^2LO,\,B}^{ijkl}
\nonumber\\
&&\hspace{0.25cm}
-\bigg[\dot{\bm{\mu}}_{1\,i}\dot{\bm{\mu}}_{1\,j}\bm{\mu}_{2\,k}\bm{\mu}_{2\,l}+\bm{\mu}_{1\,i}\bm{\mu}_{1\,j}\dot{\bm{\mu}}_{2\,k}\dot{\bm{\mu}}_{2\,l}
+\frac{1}{4}\left(\bm{\mu}_{1\,i}\dot{\bm{\mu}}_{1\,j}-\dot{\bm{\mu}}_{1\,i}\bm{\mu}_{1\,j}\right)
\left(\bm{\mu}_{2\,k}\dot{\bm{\mu}}_{2\,l}-\dot{\bm{\mu}}_{2\,k}\bm{\mu}_{2\,l}\right)\bigg]\lambda_{N^2LO,\,B}^{ijlk}\bigg\} ,
\end{eqnarray}
where $\dot{\bm{\mu}}=i\left[\bm{\mu},\hat{h}_0\right]$, and 
\begin{eqnarray}
\lambda_{N^2LO,\,B}^{ijkl} &=& \bigg\{(d-6)(d-4)\frac{\bm{k}^i\bm{k}^j\bm{k}^k\bm{k}^l}{\bm{k}^4}+\left(\delta^{ij}\delta^{kl}+\delta^{il}\delta^{jk} -2\delta^{ik}\delta^{jl}\right)
\nonumber\\
&&
+\frac{1}{\bm{k}^2}\left[ (d-4) (\bm{k}^i\bm{k}^j\delta^{kl}+\bm{k}^i\bm{k}^l\delta^{jk}+\bm{k}^k\bm{k}^l\delta^{ij}+\bm{k}^j\bm{k}^k\delta^{il})
+2(d-2)(\bm{k}^i\bm{k}^k\delta^{jl}+\bm{k}^j\bm{k}^l\delta^{ik})\right]\bigg\} \,.
\end{eqnarray}

For electric-magnetic dipole interactions the LO contribution vanishes, the following two terms read
\begin{eqnarray}
\widetilde{V}^{\text{1loop}}_{NLO,M} &=& 
-\frac{i\pi^2\alpha^2 A_{2}(\bm{k}^2)}{(d-2)m^3}
\left[\bm{L}_1^i\left(\bm{\mu}_2\times\bm{\mu}_2\right)^j + \bm{L}_2^i\left(\bm{\mu}_1\times\bm{\mu}_1\right)^j\right] \left[(d-3)\bm{k}^i\bm{k}^j+\bm{k}^2\delta^{ij}\right],
\\
\widetilde{V}^{\text{1loop}}_{N^2LO,M} &=& 
\frac{i2\pi^2\alpha^2A_{3/2}(\bm{k}^2)}{m^3\bm{k}^2}
\left[\left(\bm{\mu}_1\cdot\dot{\bm{\mu}}_1+\bm{\mu}_2\cdot\dot{\bm{\mu}}_2\right)\delta^{ij} - \bm{\mu}^i_1\dot{\bm{\mu}}^j_1 - \bm{\mu}^i_2\dot{\bm{\mu}}^j_2\right]
\left[(d-4)\bm{k}^i\bm{k}^j + \bm{k}^2\delta^{ij}\right].
\end{eqnarray}

Next we provide the one-loop matching contributions from pNRQED$^{\prime}$ to the WEFT potential of Sec.~\ref{h1vdeft} generalized to any state of the hydrogen atoms. The first contribution corresponds to the two dipole potential exchange (diagram~{\it(a)} of Fig.~\ref{h1m2})
\begin{equation}
\widetilde{W}^{(a)}_{y}=\sideset{}{'}\sum_{m_1,m_2}p_y(n_1,m_1)^{ij}p_y(n_2,m_2)^{kl}\lambda^{ijkl}_{(a)}\frac{ A_2(\bm{k}^2)}{\Delta E_{n_1m_2}+\Delta E_{n_2m_2}}\,, \qquad y=E,\,B\,,
\label{genh1lope}
\end{equation}
with
\begin{equation}
\begin{split}
\lambda^{ijkl}_{(a)}  = \frac{1}{16 d(d-2)} &\left[(d^2-4d+3)\bm{k}^i\bm{k}^j\bm{k}^k\bm{k}^l+\bm{k}^4\left(\delta^{ij}\delta^{kl}+\delta^{il}\delta^{jk}+\delta^{ik}\delta^{jl}\right) \right.
\\
&\left. +(d-1)\bm{k}^2(\bm{k}^i\bm{k}^j\delta^{kl}+\bm{k}^i\bm{k}^l\delta^{jk}+\bm{k}^k\bm{k}^l\delta^{ij}+\bm{k}^j\bm{k}^k\delta^{il})-(d+1)\bm{k}^2(\bm{k}^i\bm{k}^k\delta^{jl}+\bm{k}^j\bm{k}^l\delta^{ik})\right] .
\end{split}
\end{equation}

The second type of diagrams corresponds to the exchange of a LO and a NLO dipole potential (diagram~{\it(b)} of Fig.~\ref{h1m2})
\begin{equation}
\widetilde{W}^{(b)}_{y}=\sideset{}{'}\sum_{m_1,m_2}p_y(n_1,m_1)^{ij}p_y(n_2,m_2)^{kl}\lambda^{ijkl}_{(b)}\frac{A_2(\bm{k}^2)\D E_{n_1m_1}\D E_{n_2m_2}}{\D E_{n_1m_1}+\D E_{n_2m_2}}\,, \qquad y=E,\,B\,,
\label{genh1nlope}
\end{equation}
with  
\begin{equation}
\begin{split}
\lambda^{ijkl}_{(b)}  =  \frac{1}{8(d-2)} &\bigg[(d-5)(d-3)\frac{\bm{k}^i\bm{k}^j\bm{k}^k\bm{k}^l}{\bm{k}^2}+(d-1)(\bm{k}^i\bm{k}^k\delta^{jl}
+\bm{k}^j\bm{k}^l\delta^{ik}) 
\\
&+\bm{k}^2\left(\delta^{ij}\delta^{kl}+\delta^{il}\delta^{jk}-3\delta^{ik}\delta^{jl}\right)
+(d-3)(\bm{k}^i\bm{k}^j\delta^{kl}+\bm{k}^i\bm{k}^l\delta^{jk}+\bm{k}^k\bm{k}^l\delta^{ij}+\bm{k}^j\bm{k}^k\delta^{il})\bigg] .
\end{split}
\end{equation}
The velocity $\bm{v}$ (or $\dot{\bm{\mu}}$ in the magnetic case) that appears in the NLO order pNRQED$^{\prime}$ potential, involved in the calculation of diagram {\it (b)} of Fig.~\ref{h1m2}, does not appear in Eq.~\eqref{genh1nlope} for we have used $\langle n|\bm{v}|m\rangle = i\D E_{nm} \langle n|\bm{x}|m\rangle$ (or the equivalent for the magnetic case). 

Since the LO electric-magnetic dipole potential is proportional to the energy, the corresponding two LO potential exchange diagrams give a contribution more similar to Eq.~\eqref{genh1nlope} than Eq.~\eqref{genh1lope}:
\begin{eqnarray}
\widetilde{W}^{(a)}_{M} &=& \sideset{}{'}\sum_{m_1,m_1}\left[p_E(n_1,m_2)^{ij}p_B(n_2,m_2)^{kl}+p_B(n_1,m_1)^{ij}p_E(n_2,m_2)^{kl}\right]
\nonumber\\
&& \hspace{3cm}
\times \left[ (d-3)\bm{k}^r\bm{k}^s+\bm{k}^2\delta^{rs}\right] \epsilon^{rik}\epsilon^{sjl}\frac{A_2(\bm{k}^2)\D E_{n_1m_1}\D E_{n_2m_2}}{4(d-2)\left(\D E_{n_1m_1}+\D E_{n_2m_2}\right)}.
\end{eqnarray}

The last type of diagrams are formed by a potential interaction and an ultrasoft photon (diagrams~{\it(d)} and {\it(e)} of Fig.~\ref{h1m2}). The potential can be written as
\begin{eqnarray}
\widetilde{W}^{(d+e)}_{y} = -\sideset{}{'}\sum_{m_1,m_2}p_y(n_1,m_1)^{ij}p_y(n_2,m_2)^{kl}
&& \left[\lambda^{ijkl}_{(d+e),\,y}\frac{\D E^2_{n_1m_1} J(\D E_{n_1m_1})+\D E^2_{n_2m_2} J(\D E_{n_2m_2})}{\D E_{n_1m_1}+\D E_{n_2m_2}} \right.
\nonumber \\
&& \hspace{-1cm}
\left.-\lambda^{ijlk}_{(d+e),\,y}\frac{\D E^2_{n_1m_1} J(\D E_{n_1m_1})-\D E^2_{n_2m_2} J(\D E_{n_2m_2})}{\D E_{n_1m_1}-\D E_{n_2m_2}}\right], \; y=E,\,B \,,
\end{eqnarray} 
with
\begin{eqnarray}
\lambda^{ijkl}_{(d+e),\,E}&=&\frac{2(d-2)}{d-1}\frac{\bm{k}^j\bm{k}^l\delta^{ik}}{\bm{k}^2}\,,
\\
\lambda^{ijkl}_{(d+e),\,B}&=&\frac{2(d-2)}{d-1}\frac{(\bm{k}^j\bm{k}^l-\bm{k}^2\delta^{jl})\delta^{ik}}{\bm{k}^2}\,.
\end{eqnarray}
The equivalent contribution with two electric dipoles and two magnetic dipoles vanishes.

\subsection{Long-range regime}
The contributions to the WEFT$^{\prime}$ potential of Sec.~\ref{h2vdwp} from the two-photon exchange diagrams in pNRQED (see Fig.~\ref{h21}) generalized to any hydrogen atom state read
\begin{align}
(\widetilde{W}')^{\text{1loop}}_{LO,\,y} = -\sideset{}{'}\sum_{m_1,m_2}p_y(n_1,m_1)^{ij}p_y(n_2,m_2)^{kl}
&\left[\frac{\s_{LO,y}^{ijkl}}{\D E_{n_1m_1}+\D E_{n_2m_2}}\left(\D E^2_{n_1m_1} J(\D E_{n_1m_1}) +\D E^2_{n_2m_2}J(\D E_{n_2m_2})\right) \right.
\nonumber \\
&\hspace{-2cm}
\left. -\frac{\s_{LO,y}^{ijlk}}{\D E_{n_1m_1}-\D E_{n_2m_2}}\left(\D E^2_{n_1m_1} J(\D E_{n_1m_1})-\D E^2_{n_2m_2} J(\D E_{n_2m_2})\right) \right],
\\
(\widetilde{W}')^{\text{1loop}}_{NLO,\,y} =-\sideset{}{'}\sum_{m_1,m_2}p_y(n_1,m_1)^{ij}p_y(n_2,m_2)^{kl}
&\left[\frac{\s_{NLO,y}^{ijkl}}{\D E_{n_1m_1}+\D E_{n_2m_2}}\left(J(\D E_{n_1m_1})+J(\D E_{n_2m_2})\right)\right.
\nonumber \\
&\hspace{-2cm}
\left.-\frac{\s_{NLO,y}^{ijlk}}{\D E_{n_1m_1}-\D E_{n_2m_2}}\left(J(\D E_{n_1m_1})-J(\D E_{n_2m_2})\right) \right],
\\
(\widetilde{W}')^{\text{1loop}}_{N^2LO,\,y} = -\sideset{}{'}\sum_{m_1,m_2}\frac{p_y(n_1,m_1)^{ij}p_y(n_2,m_2)^{kl}}{\D E^2_{n_1m_1}\D E^2_{n_2m_2}}
&\left[\frac{\s_{N^2LO,y}^{ijkl}}{\D E_{n_1m_1}+\D E_{n_2m_2}}\left(\D E^2_{n_2m_2} J(\D E_{n_1m_1}) +\D E^2_{n_1m_1} J(\D E_{n_2m_2})\right) \right.
\nonumber \\
& \hspace{-2cm}
\left. -\frac{\s_{N^2LO,y}^{ijlk}}{\D E_{n_1m_1}-\D E_{n_2m_2}}\left(\D E^2_{n_2m_2} J(\D E_{n_1m_1})-\D E^2_{n_1m_1} J(\D E_{n_2m_2})\right) \right] ,
\end{align}
where $y=E,\,B,\,M$. 
In the case of electric dipole interactions we have 
\begin{eqnarray}
\s_{LO,\,E}^{ijkl}&=&-\frac{1}{2(d-1)}\left[\delta^{ij}\delta^{kl}+\delta^{il}\delta^{jk}+(d^2-6d+6)\delta^{ik}\delta^{jl}\right], 
\\
\s_{NLO,\,E}^{ijkl}&=&\frac{1}{24}\left[2\left(\bm{k}^i\bm{k}^j\delta^{kl}+\bm{k}^k\bm{k}^l\delta^{ij}
+\bm{k}^i\bm{k}^l\delta^{jk}+\bm{k}^j\bm{k}^k\delta^{il}\right)+4(d-4)\left(\bm{k}^i\bm{k}^k\delta^{jl}+\bm{k}^j\bm{k}^l\delta^{ik}\right)\right. 
\nonumber \\
&&\hspace{1cm}
\left.+\bm{k}^2\left(\delta^{ij}\delta^{kl}+\delta^{il}\delta^{jk}+(d^2-10d+22)\delta^{ik}\delta^{jl}\right)\right], 
\\
\s_{N^2LO,\,E}^{ijkl}&=&-\frac{1}{480}\left[(d-3)\bm{k}^4\left(\delta^{ij}\delta^{kl}+\delta^{il}\delta^{jk}+(d^2-14d+46)\delta^{ik}\delta^{jl}\right)+8(d-3)\bm{k}^i\bm{k}^j\bm{k}^k\bm{k}^l\right. 
\nonumber \\
&&\hspace{1cm}
\left.+4(d-3)\bm{k}^2\left(\bm{k}^i\bm{k}^j\delta^{kl}+\bm{k}^k\bm{k}^l\delta^{ij}+\bm{k}^i\bm{k}^l\delta^{jk}+\bm{k}^j\bm{k}^k\delta^{il}\right)\right. 
\nonumber \\
&&\hspace{1cm}
\left.\left. +6(d^2-9d+18)\bm{k}^2(\bm{k}^i\bm{k}^k\delta^{jl}+\bm{k}^j\bm{k}^l\delta^{ik}\right)\right],
\end{eqnarray}
and, when the interaction is mediated by magnetic dipoles,
\begin{eqnarray}
\s_{LO,\,B}^{ijkl}&=&-\frac{1}{2(d-1)}\left[\delta^{ij}\delta^{kl}+\delta^{il}\delta^{jk}+(d^2-2d-2)\delta^{ik}\delta^{jl}\right], 
\\
\s_{NLO,\,B}^{ijkl}&=&\s_{NLO,\,E}^{ijkl} \,,
\\
\s_{N^2LO,\,B}^{ijkl}&=&\s_{N^2LO,\,E}^{ijkl} \,.
\end{eqnarray}
For electric dipoles interacting with magnetic dipoles the instantaneous dipole factor $p_M(n_1,m_1)^{ij}p_M(n_2,m_2)^{kl}$ should be understood as $p_E(n_1,m_1)^{ij}p_B(n_2,m_2)^{kl}$ $+$ $p_B(n_1,m_1)^{ij}p_E(n_2,m_2)^{kl}$, and
\begin{eqnarray}
\s_{LO,\,M}^{ijkl}&=&-\frac{1}{2}\left(\delta^{ij}\delta^{kl}-\delta^{il}\delta^{jk}\right) \,,
\\
\s_{NLO,\,M}^{ijkl}&=&\frac{d-3}{24}\epsilon^{rik}\epsilon^{sjl}\left(\bm{k}^2\delta^{rs}+2\bm{k}^r\bm{k}^s\right) \,,
\\
\s_{N^2LO,\,M}^{ijkl}&=&-\frac{(d-5)(d-3)\bm{k}^2}{480}\epsilon^{rik}\epsilon^{sjl}\left(\bm{k}^2\delta^{rs}+4\bm{k}^r\bm{k}^s\right).
\end{eqnarray}

The contribution to the WEFT potential from the one-loop diagram with seagull vertices in Fig.~\ref{h22} for an arbitrary electric polarization tensor [see Eq.~\eqref{pol0}] reads 
\begin{align}
\widetilde{W}_E^{\text{seg}} = -\frac{\pi^2 \alpha_{n_1}^{ij} \alpha_{n_2}^{kl}}{16(d-1)(d+1)}
&\left[2d(d-2)\bm{k}^i\bm{k}^j\bm{k}^k\bm{k}^l-(d+4)\bm{k}^2\left(\bm{k}^i\bm{k}^l\delta^{kj}+\bm{k}^j\bm{k}^k \delta^{il}+\bm{k}^i\bm{k}^k\delta^{jl} +\bm{k}^j\bm{k}^l\delta^{ik}\right)
\right.
\nonumber \\
&\left.
+2d\,\bm{k}^2\left(\bm{k}^i\bm{k}^j\delta^{k l}+\bm{k}^k\bm{k}^l\delta^{ij}\right)
+\bm{k}^4\left(2\delta^{ij}\delta^{kl}+7(\delta^{jk}\delta^{il}+\delta^{ik}\delta^{jl})\right)\right]A_{3/2}(\bm{k}^2)\,.
\end{align}
Replacing the electric polarizabilities with the magnetic ones yields $\widetilde{W}_B^{\text{seg}}$. For the case with electric-magnetic polarizability couplings, we obtain
\begin{equation}
\widetilde{W}_M^{\text{seg}} = 
\frac{\pi^2}{16(d-1)(d+1)} \left(\alpha_{n_1}^{ij} \beta_{n_2}^{kl} + \beta_{n_1}^{ij} \alpha_{n_2}^{kl}\right)
\left(\epsilon^{rik}\epsilon^{sjl}+\epsilon^{ril}\epsilon^{sjk}\right) \bm{k}^2 \left(\bm{k}^2\delta^{rs} +d\,\bm{k}^r\bm{k}^s\right) A_{3/2}(\bm{k}^2)\,.
\end{equation}

\subsection{Intermediate-range regime}

The contribution of the one-loop pNRQED diagrams with two-photon exchanges of Fig.~\ref{h3m} to the WEFT potential for an arbitrary hydrogen atom state is 
\begin{eqnarray}
&&\widetilde{W}^{\text{1loop}}_y=-\sideset{}{'}\sum_{m_1,m_2}p_y(n_1,m_1)^{ij}p_y(n_2,m_2)^{kl}\Bigg\{\left(\frac{\ka^{ijlk}_{y02}+\ka^{ijlk}_{y22}\D E^2_{n_1m_1}}{\D E_{n_1m_1}-\D E_{n_2m_2}}
-\frac{\ka^{ijkl}_{y02}+\ka^{ijkl}_{y22}\D E^2_{n_1m_1}}{\D E_{n_1m_1}+\D E_{n_2m_2}}\right)J(\D E_{n_1m_1}) 
\nonumber\\
&&
-\left(\frac{\ka^{ijlk}_{y02}+\ka^{ijlk}_{y22}\D E^2_{n_2m_2}}{\D E_{n_1m_1}-\D E_{n_2m_2}}+\frac{\ka^{ijkl}_{y02}+\ka^{ijkl}_{y22}\D E^2_{n_2m_2}}{\D E_{n_1m_1}+\D E_{n_2m_2}}\right)J(\D E_{n_2m_2}) 
\nonumber\\
&&+\left(\frac{\ka^{ijlk}_{y00}+\ka^{ijlk}_{y20}\D E^2_{n_1m_1}+\ka^{ijlk}_{y40}\D E^4_{n_1m_1}}{\D E_{n_1m_1}-\D E_{n_2m_2}}-\frac{\ka^{ijkl}_{y00}+\ka^{ijkl}_{y20}\D E^2_{n_1m_1}
+\ka^{ijkl}_{y40}\D E^4_{n_1m_1}}{\D E_{n_1m_1}+\D E_{n_2m_2}}\right)K(\bm{k}^2,\,\D E_{n_1m_1}) 
\nonumber\\
&&-\left(\frac{\ka^{ijlk}_{y00}+\ka^{ijlk}_{y20}\D E^2_{n_2m_2}+\ka^{ijlk}_{y40}\D E^4_{n_2m_2}}{\D E_{n_1m_1}-\D E_{n_2m_2}}+\frac{\ka^{ijkl}_{y00}+\ka^{ijkl}_{y20}\D E^2_{n_2m_2}
+\ka^{ijkl}_{y40}\D E^4_{n_2m_2}}{\D E_{n_1m_1}+\D E_{n_2m_2}}\right)K(\bm{k}^2,\,\D E_{n_2m_2}) 
\nonumber\\
&&-\frac{1}{4}\bigg[\left(\ka^{ijkl}_{y40}-\ka^{ijlk}_{y40}\right)\left(\frac{\bm{k}^2}{4(d-1)}+\D E^2_{n_1m_1}+\D E^2_{n_2m_2}\right) 
\nonumber \\
&&\hspace{0.5cm}
-\left(\ka^{ijkl}_{y40}+\ka^{ijlk}_{y40}\right)\D E_{n_1m_1}\D E_{n_2m_2} +\left(\ka^{ijkl}_{y20}-\ka^{ijlk}_{y20}\right)\bigg]A_{3/2}(\bm{k}^2)\Bigg\},  \qquad\qquad\qquad\quad y=E,\,B,\,M\,.
\end{eqnarray}
The momentum-transfer-dependent coefficients read, for electric dipole interactions
\begin{eqnarray}
\ka_{E40}^{ijkl} &=& \frac{1}{d(d-2)}\bigg\{\frac{3}{\bm{k}^4} \bm{k}^i \bm{k}^j\bm{k}^k\bm{k}^l
+\frac{1}{\bm{k}^2}\left[(d-1)\left(\bm{k}^i \bm{k}^k \delta^{jl}+\bm{k}^j\bm{k}^l \delta^{ik}\right)
-\bm{k}^i\bm{k}^j\delta^{k l}-\bm{k}^k\bm{k}^l\delta^{ij} -\bm{k}^i\bm{k}^l\delta^{kj}-\bm{k}^j\bm{k}^k \delta^{il}\right]
\nonumber\\
&&\hspace{1.5cm}
+\left[\delta^{ij}\delta^{kl}+\delta^{il}\delta^{kj}+\left(d^2-4d+1\right) \delta^{ik}\delta^{jl}\right]\bigg\},
\\
\ka_{E20}^{ijkl} &=& \frac{1}{4d(d-2)}\bigg\{\frac{2(d-3)}{\bm{k}^2} \bm{k}^i \bm{k}^j\bm{k}^k\bm{k}^l-(d-2)\left(\bm{k}^i\bm{k}^j\delta^{k l}
+\bm{k}^k\bm{k}^l\delta^{ij}+\bm{k}^i\bm{k}^l\delta^{kj}+\bm{k}^j\bm{k}^k \delta^{il}\right) 
\nonumber \\
&&\hspace{1.5cm}
-(d^2-2d-2)\left(\bm{k}^i \bm{k}^k \delta^{jl}+\bm{k}^j\bm{k}^l \delta^{ik}\right)-2\bm{k}^2\left[\delta^{ij}\delta^{kl}+\delta^{il}\delta^{kj}-\left(d-1\right) \delta^{ik}\delta^{jl}\right]\bigg\}, 
\\
\ka_{E00}^{ijkl}&=&\frac{1}{16d(d-2)}\left[(d-3)(d-1)\bm{k}^i\bm{k}^j\bm{k}^k\bm{k}^l
+(d-1)\bm{k}^2\left(\bm{k}^i\bm{k}^j\delta^{k l}+\bm{k}^k\bm{k}^l\delta^{ij}+\bm{k}^i\bm{k}^l\delta^{kj} +\bm{k}^j\bm{k}^k \delta^{il}\right)\right. 
\nonumber \\
&&\hspace{1.5cm}
\left.-(d+1)\bm{k}^2\left(\bm{k}^i \bm{k}^k \delta^{jl}+\bm{k}^j\bm{k}^l \delta^{ik}\right)
+\bm{k}^4\left(\delta^{ij}\delta^{kl}+\delta^{il}\delta^{kj}+\delta^{ik}\delta^{jl}\right)\right], 
\\
\ka_{E22}^{ijkl}&=&\frac{1}{2d(d-2)}\bigg\{\frac{3(3-d)}{\bm{k}^4}\bm{k}^i\bm{k}^j\bm{k}^k\bm{k}^l
+\frac{d-3}{\bm{k}^2}\left[\bm{k}^i\bm{k}^j\delta^{k l}
+\bm{k}^k\bm{k}^l\delta^{ij}+\bm{k}^i\bm{k}^l\delta^{kj}+\bm{k}^j\bm{k}^k \delta^{il}- (d-1)\left(\bm{k}^i \bm{k}^k \delta^{jl}+\bm{k}^j\bm{k}^l \delta^{ik}\right)\right]
\nonumber \\
&&\hspace{1.5cm}
+\frac{2d-3}{d-1}\left(\delta^{ij}\delta^{kl}+\delta^{il}\delta^{kj}\right)-\frac{2d^2-4d+3}{d-1}\delta^{ik}\delta^{jl}\bigg\}, 
\\
\ka_{E02}^{ijkl}&=& \frac{1}{8d(d-2)}\bigg[\frac{(3-d)(d-1)}{\bm{k}^2}\bm{k}^i\bm{k}^j\bm{k}^k\bm{k}^l
-(d-1)\left(\bm{k}^i\bm{k}^j\delta^{k l}+\bm{k}^k\bm{k}^l\delta^{ij}+\bm{k}^i\bm{k}^l\delta^{kj}+\bm{k}^j\bm{k}^k\delta^{il}\right)
\nonumber \\
&&\hspace{1.5cm}
+(d+1)\left(\bm{k}^i \bm{k}^k \delta^{jl}+\bm{k}^j\bm{k}^l \delta^{ik}\right)-\bm{k}^2\left(\delta^{ij}\delta^{kl}
+\delta^{il}\delta^{kj}+\delta^{ik}\delta^{jl}\right)\bigg],
\end{eqnarray}
and for magnetic dipole interactions
\begin{eqnarray}
\ka_{B40}^{ijkl}&=&\ka_{E40}^{ijkl}\,,\\
\ka_{B20}^{ijkl}&=&\ka_{E20}^{ijkl}\,,\\
\ka_{B00}^{ijkl}&=&\ka_{E00}^{ijkl}\,,\\
\ka_{B22}^{ijkl}&=&\ka_{E22}^{ijkl}+\frac{2(d-2)}{(d-1)}\delta^{ik}\delta^{jl}\,, \\
\ka_{B02}^{ijkl}&=&\ka_{E02}^{ijkl}\,.
\end{eqnarray}
For electric-magnetic dipole interactions the instantaneous dipole factor $p_M(n_1,m_1)^{ij}p_M(n_2,m_2)^{kl}$ should be understood as $p_E(n_1,m_1)^{ij}p_B(n_2,m_2)^{kl} + p_B(n_1,m_1)^{ij}p_E(n_2,m_2)^{kl}$, and
\begin{eqnarray}
\ka_{M40}^{ijkl}&=&\frac{\epsilon^{rik}\epsilon^{sjl}}{(d-2)\bm{k}^2}\left(-\bm{k}^r\bm{k}^s+\bm{k}^2\delta^{rs}\right),\\
\ka_{M20}^{ijkl}&=&-\frac{\epsilon^{rik}\epsilon^{sjl}}{4(d-2)}\left[(d-3)\bm{k}^r\bm{k}^s + \bm{k}^2\delta^{rs}\right],\\
\ka_{M00}^{ijkl}&=&0\,,\\
\ka_{M22}^{ijkl}&=&\frac{\epsilon^{rik}\epsilon^{sjl}}{2(d-2)\bm{k}^2}\left[(d-3)\bm{k}^r\bm{k}^s + \bm{k}^2\delta^{rs}\right],\\
\ka_{M02}^{ijkl}&=&0\,.
\end{eqnarray}

\section{Sums over intermediate states}
\label{sumst}
In this appendix, we present some cases in which the sum over states can be explicitly performed. First, using $\displaystyle \mathbbm{1}=\sum_m |m\rangle \langle m|$, we have 
\begin{equation}
\sum_m \langle n |\bm{x}^i| m\rangle\langle m|\bm{x}^j | n\rangle = \langle n |\bm{x}^i\bm{x}^j | n\rangle = \langle n |(\bm{x}^i)^2| n\rangle\delta^{ij},
\end{equation}
where in the last step we have made use of the reflection symmetry $\bm{x}^i\rightarrow-\bm{x}^i$ for $i=1,\,2,\,3$. Everywhere in this appendix the index $i$ is understood as not summed.

Let $\hat{h}_0 = -\bm{\nabla}^2_{\bm{x}}/(2m) - \alpha/|\bm{x}|$, then it holds that 
\begin{eqnarray}
\sum_m \langle n |\bm{x}^i| m\rangle\D E_{nm} \langle m|\bm{x}^i | n\rangle &=& 
\langle n |\bm{x}^i(E_n-\hat{h}_0)\bm{x}^i | n\rangle = 
\frac{1}{2}\langle n |\left[\bm{x}^i,E_n-\hat{h}_0\right]\bm{x}^i+\bm{x}^i\left[E_n-\hat{h}_0,\bm{x}^i\right]| n\rangle 
\nonumber\\
&=& \frac{i}{2m}\langle n |\left[\bm{x}^i,\bm{p}^i\right]|n\rangle=-\frac{1}{2m}\,,
\end{eqnarray}
and for the case $i\neq j$
\begin{eqnarray}
\sum_m \langle n |\bm{x}^i| m\rangle \D E_{nm}\langle m|\bm{x}^j | n\rangle &=& \langle n |\bm{x}^i(E_n-\hat{h}_0)\bm{x}^j | n\rangle = 
\frac{1}{2}\langle n |\left[\bm{x}^i,E_n-\hat{h}_0\right]\bm{x}^j+\bm{x}^i\left[E_n-\hat{h}_0,\bm{x}^j\right]| n\rangle 
\nonumber\\
&=& \frac{i}{2m}\langle n |\bm{x}^i\bm{p}^j-\bm{p}^i\bm{x}^j|n\rangle = \frac{i}{2m}\epsilon^{ijk}\langle n|\bm{L}^k|n\rangle\,.
\end{eqnarray}
Finally, when the sum over the states contains $\D E_{nm}^2$, it holds that 
\begin{eqnarray}
\sum_m \langle n |\bm{x}^i| m\rangle\D E_{nm}^2\langle m|\bm{x}^j | n\rangle &=& \langle n |\bm{x}^i(E_n-\hat{h}_0)^2\bm{x}^j | n\rangle 
= \langle n |\left[\bm{x}^i,E_n-\hat{h}_0\right]\left[E_n-\hat{h}_0,\bm{x}^j\right]| n\rangle 
\nonumber\\
&=& \frac{1}{m^2}\langle n |\bm{p}^i\bm{p}^j|n\rangle = \frac{1}{m^2}\langle n |(\bm{p}^i)^2|n\rangle\delta^{ij}.
\end{eqnarray}

\section{Fourier Transforms}
\label{foutra}
To evaluate the van der Waals potentials in position space we have encountered the following Fourier transforms
\begin{eqnarray}
I_{n,i_1,\dots i_L}\left(\bm{R}\right)&=&\int\frac{d^3k}{(2\pi)^3}\, e^{i\bm{k}\cdot\bm{R}} \, k^n \, \hat{\bm{k}}^{i_1}\dots \hat{\bm{k}}^{i_L}\,, 
\label{cft}\\
H_{n,i_1,\dots i_L}\left(\bm{R}\right)&=&\int\frac{d^3k}{(2\pi)^3}\, e^{i\bm{k}\cdot\bm{R}} \, k^n \, \log k^2 \, \hat{\bm{k}}^{i_1}\dots \hat{\bm{k}}^{i_L}\,,
\label{qft}
\end{eqnarray}
where $\hat{\bm{k}}^{i}=\bm{k}^{i}/k$, and $k$ is the modulus of $\bm{k}$. 
The product of unit vectors can be decomposed into a sum of spherical harmonics with angular momentum up to the total number of unit vectors: 
\begin{equation}
\hat{\bm{k}}^{i_1}\dots \hat{\bm{k}}^{i_L}= \sum^{L}_{l=0}\sum^{l}_{m=-l}C^{lm}_{i_1\dots i_L}Y^m_l(\hat{\bm{k}})\,.
\label{mtdec}
\end{equation}
Due to parity, $C^{lm}_{i_1\dots i_L}$ vanishes for even (odd) values of $l$ if the number of unit vectors is odd (even). After substituting Eq.~\eqref{mtdec} in Eq.~\eqref{cft} we obtain
\begin{eqnarray}
I_{n,i_1,\dots i_L}\left(\bm{R}\right)&=&\sum^{L}_{l=0}\sum^{l}_{m=-l}C^{lm}_{i_1\dots i_L}\int\frac{d^3k}{(2\pi)^3} \, e^{i \bm{k}\cdot \bm{R}} \, k^n \, Y^m_l(\hat{\bm{k}})\,.
\end{eqnarray}
Using the Rayleigh expansion of the exponential, the addition theorem and orthogonality of the spherical harmonics we arrive at 
\begin{eqnarray}
I_{n,i_1,\dots i_L}\left(\bm{R}\right)&=&\sum^{L}_{l=0}\sum^{l}_{m=-l}C^{lm}_{i_1\dots i_L} \, Y^m_l(\hat{\bm{R}}) \, I^R_{nl}\,,
\label{F5}
\end{eqnarray}
where $I^R_{nl}$ is defined as
\begin{eqnarray}
I^R_{nl}&=&\frac{i^l}{2\pi^2}\int^{\infty}_0 dk \, k^{n+2} \, j_l( R k)\,,
\end{eqnarray}
with $j_l$ being the spherical Bessel functions and $R$ the modulus of $\bm{R}$. 

It may be convenient to rewrite Eq.~\eqref{F5} as 
\begin{equation}
I_{n,i_1,\dots i_L}\left(\bm{R}\right) = \sum^{L}_{l=0}(\hat{\bm{R}}^{i_1}\dots \hat{\bm{R}}^{i_L})_l I^R_{nl}\,,  
\label{cft2} 
\end{equation}
with
\begin{equation}
(\hat{\bm{R}}^{i_1}\dots \hat{\bm{R}}^{i_L})_l=\sum^{l}_{m=-l}C^{lm}_{i_1\dots i_L}Y^m_l(\hat{\bm{R}})\,, 
\label{lfactor}
\end{equation}
which is the sum  of all the terms with angular momentum $l$. The same procedure can be used for $H_{n,i_1,\dots i_L}$ leading to a formula analogous to Eq.~\eqref{cft2} but with $I^R_{nl}$ replaced by 
\begin{equation}
H^R_{nl} = \frac{i^l}{2\pi^2}\int^{\infty}_0 dk \, k^{n+2} \, \log k^2 \, j_l( R k)\,.
\end{equation}

The coefficients $C^{lm}_{i_1\dots i_L}$ can be obtained from $\displaystyle \hat{\bm{R}}^{i_1}\dots \hat{\bm{R}}^{i_L}= \sum^{L}_{l=0}\sum^{l}_{m=-l}C^{lm}_{i_1\dots i_L}Y^m_l(\hat{\bm{R}})$ 
using the orthogonality relations of the spherical harmonics. Then the sum in Eq.~\eqref{lfactor} can be performed using the addition theorem:
\begin{equation}
(\hat{\bm{R}}^{i_1}\dots \hat{\bm{R}}^{i_L})_l=(2l+1)\int\frac{d\Omega'}{4\pi}\,\hat{\bm{R}}'^{i_1}\dots \hat{\bm{R}}'^{i_L}\,P_l(\hat{\bm{R}}'\hat{\cdot\bm{R}})\,,
\label{F9}
\end{equation}
where $P_l$ are Legendre polynomials. Equation~\eqref{F9} can be evaluated using
\begin{equation}
\int\frac{d\Omega}{4\pi}\,\hat{\bm{R}}^{i_1}\dots \hat{\bm{R}}^{i_N}=\frac{\delta_{N,\,\text{even}}}{(N+1)!!}\left(\delta^{i_1i_2}\dots\delta^{i_{N-1}i_N}+\text{permutations}\right).
\end{equation}

The problem of finding the Fourier transforms of Eqs.~\eqref{cft} and \eqref{qft} reduces then to the problem of computing the integrals $I^R_{nl}$ and $H^R_{nl}$~\cite{adkins:2013}. We have that 
\begin{eqnarray}
I^R_{nl}=i^l\frac{2^{n}}{\pi^{3/2}R^{n+3}}\frac{\Gamma\left(\frac{n+l+3}{2}\right)}{\Gamma\left(\frac{l-n}{2}\right)}\,,\qquad -(l+3)<n \; \hbox{and} \; n\neq l+2s\,,\;\forall s\in\mathbb{N}\,.
\label{ircon}
\end{eqnarray}
The  case $n=l$ can be obtained from the completion integral of the spherical Bessel functions
\begin{eqnarray}
I^R_{ll}=i^l\frac{(2l+1)!!}{R^{l}}\delta^3(\bm{R})\,.
\end{eqnarray}
This expression can be generalized to cases when $n=l+2s$, $\forall s\in\mathbb{N}$ by using the recurrence relations for the spherical Bessel functions and proceeding by induction:
\begin{eqnarray}
I^R_{(l+2s)l}=i^l(-1)^s\frac{(2s)!!(2l+2s+1)!!}{R^{l+2s}}\delta^3(\bm{R})\,,\qquad \forall s\in\mathbb{N}\,.
\end{eqnarray}

The integral $H^R_{nl}$ can be evaluated from Eq.~\eqref{ircon} by noticing that $\log k^2= (k^{2\epsilon}-1)/\epsilon$ in the limit $\epsilon \to 0$, which implies 
\begin{eqnarray}
H^R_{nl}=2\left.\frac{dI^R_{xl}}{dx}\right|_{x=n}\,.
\end{eqnarray}
In Table~\ref{fttab1} and Table~\ref{fttab2} we list the radial integrals of the Fourier transforms from $n=-2$ to $n=4$ and for $l=0$, 2, 4 for the cases without and with a $\log k^2$ respectively.

\begin{table}[ht]
\centering
\begin{tabular}{c|c|c|c} \hline\hline
  $n$  & $I_{n0}$                                 & $I_{n2}$                                     & $I_{n4}$                         \\ \hline
  $-2$ & $1/(4\pi R)$                             & $-1/(8\pi R)$                                & $3/(32\pi R)$                    \\ 
  $-1$ & $1/(2\pi^2R^2)$                          & $-1/(\pi^2R^2)$                              & $4/(3\pi^2R^2)$                  \\ 
  $0$  & $\delta^3(\bm{R})$                       & $-3/(4\pi R^3)$                              & $15/(8\pi R^3)$                  \\ 
  $1$  & $-1/(\pi^2R^4)$                          & $-4/(\pi^2R^4)$                              & $24/(\pi^2R^4)$                  \\ 
  $2$  & $-6\,\delta^3(\bm{R})/R^2$               & $-15\,\delta^3(\bm{R})/R^2$                  & $105/(4\pi R^5)$                 \\ 
  $3$  & $12/(\pi^2R^6)$                          & $24/(\pi^2R^6)$                              & $192/(\pi^2R^6)$                 \\ 
  $4$  & $120\,\delta^3(\bm{R})/R^4$              & $210\,\delta^3(\bm{R})/R^{4}$                & $945\delta^3(\bm{R})/R^{4}$      \\ \hline
\end{tabular}
\caption{Results for the radial integrals $I_{nl}$ for n=$-2, \dots, 4$ and $l=0$, 2, 4.}
\label{fttab1}
\end{table}

\begin{table}[ht]
\centering
\begin{tabular}{c|c|c|c} \hline\hline
  $n$  & $H_{n0}$                   & $H_{n2}$                   & $H_{n4}$                     \\ \hline
  $-2$ & $-\zeta/(4\pi R)$          & $(\zeta-3)/(8\pi R)$       & $-(6\zeta-25)/(65\pi R)$     \\ 
  $-1$ & $-\zeta/(2\pi^2R^2)$       & $(\zeta-3)/(\pi^2R^2)$     & $-(12\zeta-50)/(9\pi^2R^2)$  \\ 
  $0$  & $-1/(2\pi R^3)$            & $(3\zeta-8)/(4\pi R^3)$    & $-(15\zeta-61)/(8\pi R^3)$   \\ 
  $1$  & $(\zeta-3)/(\pi^2R^4)$     & $(4\zeta-6)/(\pi^2R^4)$    & $-(24\zeta-92)/(\pi^2R^4)$   \\ 
  $2$  & $3/(\pi R^5)$              & $15/(2\pi R^5)$            & $-(105\zeta-352)/(4\pi R^5)$ \\ 
  $3$  & $-(12\zeta-50)/(\pi^2R^6)$ & $-(24\zeta-92)/(\pi^2R^6)$ & $-16(12\zeta-25)/(\pi^2R^6)$ \\ 
  $4$  & $-60/(\pi R^7)$            & $-105/(\pi R^7)$           & $-945/(2\pi R^7)$            \\ \hline
\end{tabular}
\caption{Results for the radial integrals $H_{nl}$ for n=$-2, \dots, 4$ and $l=0$, 2, 4. We have defined $\zeta=2\gamma_E+\log{R^2}$.}
\label{fttab2}
\end{table}

Finally, we reproduce the partial-wave decompositions \eqref{lfactor} for two and four unit vectors~\cite{adkins:2013}:
\begin{eqnarray}
(\hat{\bm{R}}^{i}\hat{\bm{R}}^{j})_2&=&\hat{\bm{R}}^{i}\hat{\bm{R}}^{j}-\frac{1}{3}\delta^{ij} \,,
\\
(\hat{\bm{R}}^{i}\hat{\bm{R}}^{j})_0&=&\frac{1}{3}\delta^{ij} \,,
\\
(\hat{\bm{R}}^{i}\hat{\bm{R}}^{j}\hat{\bm{R}}^{k}\hat{\bm{R}}^{l})_4&=&
\hat{\bm{R}}^{i}\hat{\bm{R}}^{j}\hat{\bm{R}}^{k}\hat{\bm{R}}^{l}-\frac{1}{7}(\hat{\bm{R}}^{i}\hat{\bm{R}}^{j}\delta^{kl}+\text{permutations})+\frac{1}{35}(\delta^{ij}\delta^{kl}+\text{permutations})  \,,
\\
(\hat{\bm{R}}^{i}\hat{\bm{R}}^{j}\hat{\bm{R}}^{k}\hat{\bm{R}}^{l})_2&=&\frac{1}{7}((\hat{\bm{R}}^{i}\hat{\bm{R}}^{j})_2\delta^{kl}+\text{permutations}) \,,
\\
(\hat{\bm{R}}^{i}\hat{\bm{R}}^{j}\hat{\bm{R}}^{k}\hat{\bm{R}}^{l})_0&=&\frac{1}{15}(\delta^{ij}\delta^{kl}+\text{permutations}) \,,
\end{eqnarray}
where in the parentheses all the index permutations have to be added.

\bibliography{Waals}

\end{document}